\pdfoutput=1
\documentclass[preprint,superscriptaddress,nofootinbib]{revtex4-1}
\usepackage{hyperref}
\usepackage{bbm}
\usepackage{amsfonts}
\usepackage{mathrsfs}

\usepackage{amsmath,amssymb}

\usepackage{epsfig}
\usepackage{graphicx}               
\usepackage{url}
\usepackage{hyperref}
\usepackage{float}
\usepackage{pstricks}
\usepackage{color}

\newcommand{\nc}{\newcommand}

\nc{\beq}{\begin{equation}}
\nc{\eeq}{\end{equation}}
\nc{\beqa}{\begin{eqnarray}}
\nc{\eeqa}{\end{eqnarray}}
\nc{\bea}{\begin{eqnarray}}
\nc{\eea}{\end{eqnarray}}
\nc{\barray}{\begin{eqnarray}}
\nc{\earray}{\end{eqnarray}}
\nc{\barrayn}{\begin{eqnarray*}}
\nc{\earrayn}{\end{eqnarray*}}
\nc{\ra}{\rightarrow}
\newcommand{\lsim}{\!\mathrel{\hbox{\rlap{\lower.55ex \hbox{$\sim$}} \kern-.34em \raise.4ex \hbox{$<$}}}}
\newcommand{\gsim}{\!\mathrel{\hbox{\rlap{\lower.55ex \hbox{$\sim$}} \kern-.34em \raise.4ex \hbox{$>$}}}}
\nc{\Tr}{{\rm Tr}}
\nc{\slsh}{\slash\hspace*{-0.22cm}}
\def\eg{{\it e.g.}}
\def\ie{{\it i.e.}}
\def\be{\begin{equation}}
\def\ee{\end{equation}}
\def\bea{\begin{eqnarray}}
\def\eea{\end{eqnarray}}
\def\bit{\begin{itemize}}
\def\eit{\end{itemize}}

\nc{\infinity}{\infty}
\nc{\mc}{\mathcal}

\newcommand{\pv}{\langle\phi\rangle}

\newcommand{\neff}{N_{\rm eff}}
\newcommand*\diff{\mathop{}\!\mathrm{d}}

\newcommand{\M}{|\mathcal{M}|^2}
\newcommand{\pn}{p_{N_1}}

\newcommand{\Nt}{\tilde{N}_1}
\newcommand{\pNt}{p_{\tilde{N}_1}}
\newcommand{\eeff}{\eta_{\rm 1\, eff}}
\newcommand{\xeff}{x_{\rm 1\, eff}}

\def\ie{{\it i.e.}}
\def\eg{{\it e.g.}}

\begin{document}

\title{~\\Cosmological Imprints of Frozen-In Light Sterile Neutrinos}

\author{\vskip 0.5cm Samuel B. Roland and Bibhushan Shakya \vskip 0.05cm \textit{Michigan Center for Theoretical Physics, \\University of Michigan, Ann Arbor, MI 48109, USA \\ ~}}

\begin{abstract}
We investigate observable cosmological aspects of sterile neutrino dark matter produced via the freeze-in mechanism. The study is performed in a framework that admits many cosmologically interesting variations: high temperature production via annihilation processes from higher dimensional operators or low temperature production from decays of a scalar, with the decaying scalar in or out of equilibrium with the thermal bath, in supersymmetric or non-supersymmetric setups, thus allowing us to both extract generic properties and highlight features unique to particular variations. We find that while such sterile neutrinos are generally compatible with all cosmological constraints, interesting scenarios can arise where dark matter is cold, warm, or hot, has nontrivial momentum distributions, or provides contributions to the effective number of relativistic degrees of freedom $N_{\text{eff}}$ during Big Bang nucleosynthesis large enough to be probed by future measurements.
\end{abstract}

\preprint{MCTP-16-22}

\maketitle

\tableofcontents

\section{Introduction and Motivation}
\label{sec:introduction}

A sterile neutrino is a well-motivated and widely studied dark matter (DM) candidate. The traditional candidate, studied within the Neutrino Minimal Standard Model ($\nu$MSM) \cite{Asaka:2005an,Asaka:2005pn,Asaka:2006nq}, has a keV scale mass, where its mixing with the active neutrinos is appropriate for both producing the correct (warm) dark matter relic abundance through the Dodelson-Widrow (DW) mechanism \cite{Dodelson:1993je} and making it sufficiently long lived. However, this nonzero mixing also results in decays producing a monochromatic gamma ray line, which is constrained by X-ray measurements \cite{Boyarsky:2006fg,Boyarsky:2006ag, Boyarsky:2005us,Boyarsky:2007ay,Boyarsky:2007ge}, while the warm nature of DM from DW production disrupts small scale structure formation, which is constrained by Lyman-$\alpha$ measurements \cite{Seljak:2006qw, Asaka:2006nq, Boyarsky:2008xj}. The combination of these two constraints now rule out DW as a viable production mechanism for sterile neutrino dark matter (see, e.g. \cite{Horiuchi:2013noa} for a recent summary).

Several alternate production mechanisms that circumvent these bounds to various degrees exist in the literature \cite{Shi:1998km,Boyarsky:2008mt,Nemevsek:2012cd,Asaka:2006ek,Asaka:2006nq,Bezrukov:2009th,Patwardhan:2015kga,Shuve:2014doa,Khalil:2008kp,Lello:2014yha,Falkowski:2011xh,Biswas:2016bfo}. The Shi-Fuller mechanism \cite{Shi:1998km} produces a colder population but requires fine-tuned parameters to ensure resonant production, and might still be incompatible with structure formation \cite{Horiuchi:2015qri,Schneider:2016uqi}. Thermal freeze-out with additional interactions, followed by appropriate entropy dilution, can result in the correct relic abundance \cite{Bezrukov:2009th,Nemevsek:2012cd,Patwardhan:2015kga}, but faces strong constraints from Big Bang nucleosynthesis \cite{King:2012wg}. One mechanism that is particularly successful and employed widely is sterile neutrino dark matter production through the freeze-in mechanism \cite{Chung:1998rq, Hall:2009bx} via a feeble coupling to some particle beyond the Standard Model present in the early universe. This can be realized in several motivated frameworks: this particle could be the inflaton \cite{Shaposhnikov:2006xi}, a heavy higgs in an extended Higgs sector \cite{Petraki:2007gq,Kusenko:2006rh,McDonald:1993ex,Yaguna:2011qn,Merle:2013wta,Adulpravitchai:2014xna,Kang:2014cia}, a scalar that breaks a symmetry that the sterile neutrinos might be charged under \cite{Roland:2014vba,Roland:2015yoa,Heurtier:2016iac}, a charged scalar motivated by leptogenesis \cite{Frigerio:2014ifa}, the radion in warped extra dimension models \cite{Kadota:2007mv}, or pseudo-Dirac neutrinos \cite{Abada:2014zra}; for a recent review of various scenarios that admit freeze-in of sterile neutrino dark matter, see Ref.\,\cite{Shakya:2015xnx}. Such scenarios carry the dual virtues of a colder sterile neutrino population compared to DW as well as not relying on any mixing with the active neutrinos for production, thereby alleviating the tension with Lyman-$\alpha$  and X-ray measurements. \footnote{It should be clarified that DW is technically also a freeze-in mechanism; in this paper, freeze-in will be understood to refer to production mechanisms that do not involve active-sterile mixing.}

The phenomenological signatures of sterile neutrino dark matter from such freeze-in scenarios are in stark contrast to those from DW production. In the latter framework, the ``smoking gun" signature is a monochromatic X-ray line from the loop level decay into an active neutrino and a single photon, induced by the mixing between active and sterile neutrinos required for DW production. In the freeze-in scenario, this mixing angle can be arbitrarily small, and there is essentially no direct coupling between the sterile neutrino dark matter candidate and the Standard Model particles; hence no signals arising from such active-sterile mixing that characterize sterile neutrino dark matter from DW, such as astrophysical signatures in gamma rays or direct production in searches for neutral leptons in laboratory experiments \cite{PIENU:2011aa,Bergsma:1985is,Ruchayskiy:2011aa,Bernardi:1985ny,Bernardi:1987ek,Vaitaitis:1999wq}, are expected. The most promising observable imprints are instead of a cosmological nature: the phase space distribution of sterile neutrinos from freeze-in is distinct from that arising from DW, and can lead to possible deviations in free-streaming lengths of warm dark matter or the dark radiation content of the universe during Big Bang nucleosynthesis (BBN) or cosmic microwave background (CMB) decoupling. Although the exact properties depend on the details of the underlying model, given that such cosmological imprints offer the most direct probes of sterile neutrino dark matter from freeze-in, it is worth studying such features in greater detail in a broad framework.

The purpose of this paper is to investigate such potentially observable cosmological aspects of sterile neutrino dark matter. We perform this study in a specific model, based on Ref.\,\cite{Roland:2014vba}, which admits many cosmologically interesting variations: production can occur via annihilation processes from higher dimensional operators that are active at the highest temperatures (referred to as ultraviolet (UV) freeze-in), or from decays of a scalar, which occur at lower temperatures (infrared (IR) freeze-in); the scalar producing the dark matter population can be taken to be in or out of equilibrium with the thermal bath; moreover, both supersymmetric and non-supersymmetric setups can be considered. The framework therefore covers a diverse range of possibilities, allowing us to both extract generic properties and highlight features unique to particular variations. Similar studies have been performed in previous work in the literature \cite{Merle:2015oja,Petraki:2007gq}, but in a more constrained framework of a keV scale sterile neutrino with IR production only in a non-supersymmetric setup.

The paper is organized as follows. Section \ref{sec:model} outlines the theoretical framework and the various scenarios that we investigate in this paper.  Section \ref{sec:formalism} describes the formalism employed in our calculations, covering the topics of Boltzmann equations, entropy dilution, the various observables of interest, and the simplifying assumptions made in our formalism. Results of our calculations are presented for various scenarios in Section \ref{sec:results}. We conclude by summarizing our main results in Section \ref{sec:summary}. Details of the Boltzmann equations and related collision terms used to derive our results are presented in Appendix \ref{sec:Collision}.

\section{Theoretical Framework and Scenarios}
\label{sec:model}
We begin by outlining the theoretical framework for this paper, based on the model presented in  Ref.\,\cite{Roland:2014vba} (see also \cite{Roland:2015yoa}). The Standard Model (SM) is extended by three right-handed sterile neutrinos $N_{1,2,3}$, which are assumed to be charged under a new symmetry $U(1)'$. This symmetry is broken by the vacuum expectation value (vev) of a scalar $\phi$, which carries a $U(1)'$ charge opposite to that of the $N_i$, such that $N_i\phi$ is a $U(1)'$ and SM singlet. These charge assignments lead to no new renormalizable interactions, and the following terms appear at leading order (dimension five):
\beq
\mathcal{L} \supset \frac{y_{ij}}{M_*} L_i H N_j \phi + \frac{x_{i}}{M_*} \phi\, \phi N_i N_i\,,
\label{eq:lagrangian}
\eeq
where $M_*$ is the UV-cutoff for this theory (which we take to be the GUT scale $M_*=10^{16}$ GeV), $L_i$ is the SM lepton doublet of flavor $i$, $H$ is the SM Higgs doublet, and the $N_i$ are chosen to be in a basis where $x_i$ is diagonal. With vev insertions of both $\phi$ and the SM Higgs, these terms lead to the familiar Majorana and Dirac masses that give rise to the seesaw mechanism. In the above setup, the following masses for the active and sterile neutrino eigenstates and mixing between the two sectors are generated (indices have been suppressed):
\beq
m_{N_i}=\frac{x_i\pv^2}{M_*},~~~m_a=\frac{y^2 \langle H\rangle^2}{x M_*},~~~\sin\theta \approx \frac{y \langle H\rangle}{x \pv}.
\eeq

This setup is appealing since phenomenologically interesting (keV-GeV) masses for the sterile neutrinos are realized with $\mathcal{O}(1)$ values for the dimensionless couplings $x$ and $y$ and a high scale of new physics corresponding to $\pv\sim 1-100$ PeV (see \cite{Roland:2014vba} for details). The parameters are constrained by the seesaw requirement and cannot be completely arbitrary. We pick $N_1$ to be the sterile neutrino dark matter candidate. In this paper, the parameters are constrained as follows:
\begin{itemize}
\item $m_\phi$ and $\pv$ are taken to be free parameters.
\item Fixing the sterile neutrino masses fixes $x_i = M_* m_{N_i}/\pv^2$. Cosmological constraints require $N_{2,3}$ to decay before BBN \cite{Asaka:2005pn, Asaka:2006ek, Kusenko:2009up, Hernandez:2014fha, Vincent:2014rja}, constraining them to GeV scale or heavier masses. We fix $m_{2,3} =$(1.0 GeV, 1.1 GeV), which fixes $x_2,x_3$, unless specified otherwise. We leave $m_{N_1}$ (hence $x_1$) as a free parameter.
\item For fixed sterile neutrino masses, the $y_{ij}$ couplings are fixed by constraints on the active-sterile mixing angles. For the dark matter candidate $N_1$, its mixing with the active neutrinos needs to be heavily suppressed in order for it to be long-lived, which is accomplished by making the corresponding couplings arbitrarily small, essentially $y_{i1}\sim 0$ (which also renders the lightest active neutrino essentially massless). While such small couplings appear fine-tuned, the limit in which they vanish is technically natural since this enhances the framework by a $\mathbb{Z}_2$ symmetry for $N_1$. The remaining $y_{ij}$ are fixed by the requirements of matching the neutrino oscillation data (for $m_{2,3} =$(1.0 GeV, 1.1 GeV) and $\pv$ at the PeV scale, these couplings are $\mathcal{O}(1)$; see \cite{Roland:2014vba}).
\end{itemize}

\textit{Dark Matter Production:}\\
While the above formalism was implemented to naturally explain neutrino masses and light sterile neutrinos, it also opens possibilities for $N_1$ production in the early universe.

The first term in Eq.\,\ref{eq:lagrangian} leads to $\phi$ production via $L\,H\rightarrow N_{2,3}\,\phi$ (note that no $N_1$ is produced since $y_{i1}\sim 0$), and the second term leads to $\phi\rightarrow N_iN_i$ decays after $\phi$ obtains a vev. The relic abundance of $N_1$ produced in this manner is approximately \cite{Roland:2014vba}
\be
\Omega_{N_1} h^2\sim 0.1 \sum_{i,j} y_{ij}^2 \left(\frac{m_{N_1}}{\text{GeV}}\right)\left(\frac{1000\,T_{RH}\,M_{P}}{M_*^2}\right) Br(\phi\rightarrow N_1 N_1)
\ee
which is sensitive to the reheat temperature $T_{RH}$, at which $\phi$ production via $L\,H\rightarrow N_{2,3}\,\phi$ is assumed to begin.

If $\phi$ has additional interactions that are strong enough to keep it in equilibrium with the thermal bath in the early universe (these can, for instance, arise from the interaction terms that lead to $\phi$ obtaining a vev), two distinct production mechanisms are possible for $N_1$. At high temperatures, $\phi\,\phi\rightarrow N_1\,N_1$ (termed ultraviolet (UV) freeze-in) results in the approximate $N_1$ abundance \cite{Roland:2014vba}
\beq
\Omega_{N_1} h^2\sim 0.1\,x_1^2\left(\frac{m_{N_1}}{\rm{GeV}}\right)\left(\frac{1000\,T_{R}\,M_{Pl}}{M_*^2}\right).
\label{eq:nephidecay}
\eeq
Once $\phi$ obtains a vev, the decay process $\phi\rightarrow N_1\,N_1$ also occurs (termed infrared (IR) freeze-in) with an effective coupling $x_{\rm 1\, eff}=\frac{2\, x_1\, \pv}{M_*}$,  giving an approximate abundance \cite{Roland:2014vba}
\begin{equation}
\Omega_{N_1} h^2\sim 0.1 \left(\frac{x_{\rm 1\, eff}}{1.4\times10^{-8}}\right)^3 \left(\frac{\pv}{m_{\phi}}\right).
\end{equation}
In this case, we have assumed that the additional interactions cause $\phi$ to rapidly decay into SM radiation once it goes out of equilibrium, so that $N_1$ production occurs only while $\phi$ is in equilibrium.

\textit{Supersymmetric Extension:}\\
The above setup requires new physics that breaks the $U(1)'$ via a $\phi$ vev at high scales. Given that supersymmetry is well-motivated yet there are no signs of supersymmetry close to the weak scale, one can entertain the possibility that supersymmetry exists at a higher scale and the breaking of $U(1)'$ is tied to supersymmetry breaking. This consideration motivates a supersymmetric extension of the Lagrangian above. We introduce a chiral supermultiplet $\Phi$ with spin $(0,1/2)$ components ($\phi,\psi$) and three chiral supermultiplets ${\mathcal{N}}_i$ with components ($\tilde{N}_i,N_i$), leading to the superpotential
\beq
\label{eq:superpot}
{\mathcal{W}}\supset ~\frac{\xi_{ij}}{M_*} \mathcal{L}_i \mathcal{H}_u \,\mathcal{N}_j \Phi + \frac{\eta_i}{M_*}\mathcal{N}_i \mathcal{N}_i \Phi\Phi\,.
\eeq
This gives rise to the Lagrangian terms listed in Eq.\,\ref{eq:lagrangian} along with some other terms. In addition, the following soft terms that can appear in the Lagrangian after supersymmetry breaking are important for our discussion:
\beq
\label{eq:softterms}
{\mathcal{L}}\supset ~\xi_{ij}\frac{A_{\xi_{ij}}}{M_*} \tilde{L}_i h_u \,\tilde{N}_j \phi + \eta_i\frac{A_{\eta_i}}{M_*}\tilde{N}_i \tilde{N}_i \phi\phi\,,
\eeq
The first term leads to mixing between the sterile and standard sneutrinos, whereas the second term gives rise to the decay process $\phi\rightarrow \tilde{N}_j \tilde{N}_j$ if $m_\phi\,\textgreater\,2m_{\tilde{N}_j}$. For simplicity, we assume R-parity and take a sub-TeV Higgsino to be the LSP, which will thus account for a small fraction of dark matter.

In this supersymmetric extension, additional production channels and constraints come into play due to the presence of new interactions and superpartners, leading to qualitative differences from the non-supersymmetric setup. Of primary relevance are the fermion $\psi$ and the sterile sneutrinos $\tilde{N}_i$, which are assumed to have masses of the same scale as $\phi$ as they are all assumed to originate from supersymmetry breaking. The sterile sneutrinos decay via $\tilde{N}_{2,3}\rightarrow \tilde{H} \nu$ (with a $\phi$ vev insertion) or via their mixing with the standard sneutrinos induced by the soft term proportional to $A_\xi$ in Eq.\,\ref{eq:softterms}. The decay mechanism for the sterile sneutrino $\tilde{N}_1$ is more pertinent. If it has significant mixing with other sneutrinos, it decays through the standard sneutrino channels; however, if this mixing is significantly suppressed (this would be technically natural, corresponding to the same $\mathbb{Z}_2$ symmetry that makes $N_1$ long-lived), its decay must originate from the $\mathcal{N}_i \mathcal{N}_i \Phi\Phi$ term in the superpotential. We assume $m_{\tilde{N}_1}\textgreater \,m_\psi$, so that $\tilde{N}_1$ decays via  $\tilde{N}_1 \rightarrow \psi\,N_1$ (with a $\phi$ vev insertion), such that each $\tilde{N}_1$ decay produces one $N_1$ particle, while $\psi$ decays as $\psi\rightarrow \nu \tilde{H} N_{2,3}$. To avoid non-thermal production of the LSP at late times, we require these decays to occur before LSP decoupling (a sub-TeV $m_{\tilde{H}}$ can generally be picked to satisfy this constraint, unless extreme values of the parameters are chosen). The other choice $m_\psi\,\textgreater\,m_{\tilde{N}_1}$ requires $\tilde{N}_1$ to decay via an off-shell $\psi$ and generally has an extremely long lifetime that leads to inconsistencies, hence we do not consider it further.

Based on the above possibilities, we will divide our study into the following scenarios:

\begin{itemize}
\item Scenario I: $\phi$ in equilibrium, no supersymmetry
\item Scenario II: $\phi$ freezes in, no supersymmetry
\item Scenario III: $\phi$ in equilibrium, supersymmetry
\item Scenario IV: $\phi$ freezes in, supersymmetry
\end{itemize}

We will consider each scenario in detail in turn in Section \ref{sec:results}. Before that, we turn to a discussion of the formalism we employ to perform our studies.

\section{Formalism}
\label{sec:formalism}

All the information relevant for calculating various quantities of interest is contained in the phase space distribution of the sterile neutrinos. In this section, we describe our formalism for tracking this phase space distribution from when these particles are produced to the present era, and the subsequent calculation of the various observables of interest.

\subsection{Boltzmann Equations}
The evolution of the phase space density of particles is given by the Boltzmann equations. These take the form $L[f] = C[f]$, where the Liouville operator $L$ is

\be
L = \frac{\partial}{\partial t} - Hp\frac{\partial}{\partial p}\,
\ee
with H the Hubble parameter, and $C[f]$ is a sum of collision terms, each corresponding to an interaction. Here $f = f(\textbf{p},T)$ is the phase space density of a particle species, whose distribution is assumed to be homogenous and isotropic. We use the photon temperature $T$ to track the evolution of the phase space density. The universe is generally radiation dominated throughout the period of interest, so that
\be
H(T) = \frac{T^2}{M_0}\,, ~\text{with} ~ M_0 = \left(\frac{45 M_{Pl}^2}{4\pi^3 g_*}\right)^{1/2}\,,
\ee
where $g_*$ is the number of degrees of freedom in the bath. In some scenarios, there are heavy long-lived particles that introduce a period of matter domination, modifying the above relation; we account for such effects where necessary.

Following \cite{Merle:2015oja}, we work with the coordinates $x_i=p_i/T\,,~r=m_\phi/T$ (where $i$ denotes the particle species of interest), which leads to a simplification of the Liouville operator
\be
L = Hr\frac{\partial}{\partial r}\,
\ee
assuming $g_*$ is constant, which is a good approximation for various stages of sterile neutrino production we study in this paper.

The collision term for a particular phase space density $f_X$ and scattering process $X+i+j+\ldots \leftrightarrow a+b+\ldots$ is given by:
\be
\label{eq:collision}
C[f_X] = \frac{1}{2 E_X} \int\left(\prod_{I=i,j,\ldots} \diff\Pi_I \right) \left(\prod_{A=a,b,\ldots} \diff\Pi_A \right) (2\pi)^4  \, \delta^4\left(\Sigma p\right)\, |\mathcal{M}|^2\, \Omega(X+i+j+\ldots \leftrightarrow a+b+\ldots)\,,
\ee
with
\be
\diff\Pi_x = \frac{g_x}{(2\pi)^3} \frac{\diff^3p_x}{2 E_x}\,,
\ee
where $g_x$ counts the internal degrees of freedom of particle $x$. The factor $\Omega$ is the phase space density weight, given by
\be
\Omega(X+i+j+\ldots \leftrightarrow a+b+\ldots) = f_i f_j \ldots f_X (1\pm f_a)(1\pm f_b)\ldots - f_a f_b \ldots (1\pm f_i)(1\pm f_i)\ldots (1\pm f_X)\,,
\ee
with + for bosons and - for fermions. $|\mathcal{M}|^2$ is the squared matrix element for the scattering process of interest, averaged over initial and final states, including any symmetry factors.

Details of the Boltzmann equations and collision terms for each scenario are presented in Appendix \ref{sec:Collision}. For a detailed discussion of several subtle factors in solving the Boltzmann equations for the freeze-in of sterile neutrinos, we refer the interested reader to Ref.\,\cite{Konig:2016dzg}.


\subsection{Degrees of Freedom and Entropy Dilution}

An important aspect of calculating the abundance and momentum distribution of sterile neutrino dark matter is taking into account any changes in the effective number of degrees of freedom, $g_*$, and entropy, $S$, between dark matter production and the present epoch. Since $N_1$ is out of equilibrium from the moment of production, such changes in $S$ and $g_*$ will heat up the thermal bath without introducing any energy into the dark sector, therefore redshifting its momentum relative to the visible sector as well as diluting its abundance. There are several such major transitions:
\begin{enumerate}
\item Reduction of the supersymmetric degrees of freedom, around $T\sim$ $\pv$. Before superpartners decouple, ${g_*}_{\rm SUSY} \sim 300$ \footnote{Since the theory contains $\phi$, $\psi$, and possibly additional fields involved with $U(1)'$ breaking, the field content is presumably much larger than the MSSM, and we use  ${g_*}_{\rm SUSY} \sim 300$ as a representative value; our final results are not very sensitive to the exact choice for this number.}, which drops to ${g_*}_{\rm SM} \approx 100$.
\item Reduction of the SM degrees of freedom. This reduces ${g_*}_{\rm SM}\approx100$ above electroweak temperatures to ${g_*}_{0}=3.91$ at present.
\item Decay of the additional sterile neutrinos $N_{2,3}$.
\item Decay of the sterile sneutrino $\tilde{N}_1$. This needs to be treated separate from the rest of the supersymmetric spectrum as $\tilde{N}_1$ is long-lived and can lead to a period of matter domination before it decays.
\end{enumerate}

For simplicity, we assume that DM production, as well as $\phi,\,\psi,$ and $\tilde{N}_i$ production, take place during epochs of constant $g_*$. For a decoupled species $X$, using the fact that its momentum simply redshifts with the scale factor as $p=\frac{a_i}{a} p_i$, and that the scale factor is related to entropy by $S=g_* T^3 a^3$, we can write
\be
f_X(p,t_f) = f_X\left(\left(\frac{S_f}{S_i}\right)^{1/3}  \left(\frac{g_{*i}}{g_{*f}}\right)^{1/3} \frac{T_i}{T_f} p , t_i\right)\,,
\ee
where the subscripts $i, f$ denote initial and final values. Likewise, the number density $n_X$ and yield $Y_X = n_X/s$, where $s$ is the entropy density, scale as
\begin{align}
n_X(t_f) &= \frac{S_i}{S_f}  \frac{g_{*f}}{g_{*i}} \left(\frac{T_f}{T_i}\right)^3 n_X(t_i), \\
Y_X(t_f) &=  \frac{S_i}{S_f} Y_X(t_i)\,.
\end{align}

Calculating the entropy dilution from the decay of the heavier (GeV scale) sterile neutrinos $N_{2,3}$ is slightly involved as they thermalize, decouple while still relativistic around $\mathcal{O}(20)$ GeV \cite{Asaka:2006ek}, and decay late (just before BBN). The ratio of entropy from $N_{2,3}$ decays to the entropy in the remainder of the system, which provides the suppression factor for the dark matter relic density, is calculated to be  \cite{Scherrer:1984fd,Bezrukov:2009th,Asaka:2006ek}
\begin{equation}
S_{N23}\approx \left(1+\sum_{i=2,3} 2.95\left(\frac{2\pi^2\bar{g}_*}{45}\right)^{1/3}\left(\frac{Y_{N_i}^2 m_{N_i}^2}{M_{Pl} \Gamma_{N_i}},\right)^{2/3}\right)^{3/4}
\label{sns}
\end{equation}
where $\Gamma_{N_i}$ is the decay width of the sterile neutrino $N_i$, $\bar{g}_*$ is the average effective number of degrees of freedom during $N_{2,3}$ decay, and $Y_{N_i}$ is the yield abundance when $N_i$ decouples, given by \cite{Scherrer:1984fd,Bezrukov:2009th}
\begin{equation}
Y_{N_i}=\frac{135\, \zeta(3)}{4\pi^4g_*},
\end{equation}
where $g_*$ represents the number of degrees of freedom when $N_i$ decouples. The numerical value of $S_{N23}$ can thus be estimated by calculating the decay widths $\Gamma_{N_i}$ \cite{Bezrukov:2009th} and using the information that $N_{2,3}$ decouple around $\mathcal{O}(20)$ GeV \cite{Asaka:2006ek}. For GeV scale or heavier $N_{2,3}$, this results in $S_{N23}\lsim 30$.

If the sterile sneutrino $\tilde{N}_1$ is sufficiently long-lived and abundant that its energy density grows to be comparable to or larger than the total energy density in the thermal bath, its decays lead to a significant entropy dump into the thermal bath, significantly raising its temperature. In this scenario, the amount of entropy released from $\tilde{N}_1$ decay relative to the entropy present in the bath, and the temperature the bath is heated to from such decays, can be calculated as \cite{Kolb:1990vq}:
\begin{align}
\frac{S_f}{S_i}&\approx 1.83\, g_*^{1/4} \,\frac{m_{\tilde{N}_1}Y_{\tilde{N}_1}\tau_{\tilde{N}_1}^{1/2}}{M_{Pl}^{1/2}}\,,\\
T_{decay} &\approx 0.55\, g_*^{-1/4}\, (M_{Pl}/\tau_{\tilde{N}_1})^{1/2},\,
\end{align}
where $\tau_{\tilde{N}_1}$ is the lifetime of the sterile sneutrino.

\subsection{Observables}

The phase space distribution calculated from the above prescription can be used to calculate several observables of interest. The ones we study in this paper are as follows:

\begin{itemize}
\item The relic density $\Omega_{N_1}$, which can be expressed in terms of the distribution $f_{N_1}(x,T)$ as:
\beq
\Omega_{N_1} = \frac{n_{N_1} m_{N_1}}{\rho_c} = \frac{g_{N_1} m_{N_1} T^3}{2 \pi^2 \rho_c} \int_0^\infty \diff x ~ x^2 f_{N_1}(x,T)\,,
\eeq
where ${\rho_c}$ is the critical density.

\item $\Delta\neff$(BBN), the contribution to the effective number of relativistic degrees of freedom during BBN. This can be estimated as
\begin{eqnarray}
\Delta N_{\rm{eff}} (\text{BBN})&=&\frac{\rho_{N_1}- n m_{N_1}}{\rho_{\nu}}\nonumber\\
&=& \frac{120}{7 \pi^4} \frac{m_{N_1}}{T_{BBN}} \int_0^\infty \diff x\, x^2\left( \sqrt{ 1+ \left(\frac{x}{m_{N_1}/T_{BBN}}\right)^2 } -1 \right)f_{N_1}(x,T_{BBN}),
\end{eqnarray}
which compares the kinetic part of the sterile neutrino energy density with the energy density of a neutrino species in equilibrium at the same temperature, and we take $T_{BBN}=4$ MeV. Current measurements bound this contribution at the level of $\Delta N_{\rm{eff}}\, (\text{BBN})\,\lsim\, 0.5$ \cite{Cyburt:2015mya}, and $\mathcal{O}(0.1)$ values might be probed by future measurements. There exist stronger bounds on $\Delta N_{\rm{eff}}$ from the era of Cosmic Microwave Background (CMB) decoupling; however, these are generally less stringent for sterile neutrino dark matter as it tends to redshift and become nonrelativistic by this time \cite{Merle:2015oja}. Therefore, we only consider bounds from the BBN era in this paper.

\item Free-streaming length $\Lambda_{FS}$. This is calculated as the average distance traveled by a DM particle since the time of production:
\beq
\Lambda_{FS} = \int_{T_p}^{T_0} \frac{\langle v(T) \rangle}{a(T)} \frac{\diff t}{\diff T} \diff T \,.
\label{fseq}
\eeq
The average velocity of a DM particle is calculated using the phase space distribution function as
\beq
\langle v(T) \rangle = \frac{ \int_0^\infty \diff x ~ \frac{x^3}{\sqrt{x^2+(m_{N_1}/T)^2}} f_{N_1}(x,T) }{ \int_0^\infty \diff x ~ x^2 f_{N_1}(x,T)}\,.
\eeq
As a rough guide, we take the regimes for cold, warm, and hot dark matter to be approximately $\Lambda_{FS}\lesssim0.01$ Mpc, $0.01\lesssim \Lambda_{FS} \lesssim 0.1$ Mpc, and $0.1 ~\text{Mpc}\lesssim\Lambda_{FS}$ respectively \cite{Merle:2015oja}; we further discuss the subtleties related to using the free-streaming length as a proxy for a measure of impact on structure formation in Sec.\,\ref{ss:fsandsf}. 
\end{itemize}

\subsection{Simplifying Approximations}

We have made several simplifying assumptions and approximations in the formalism described above. In this subsection, we discuss these assumptions and their possible effects on the results discussed in this paper. 

\subsubsection{Additional Particle Content and Dynamics}
In this paper, we only focus on the ``minimal" phenomenology arising from the fields and interactions listed in Eq.\,\ref{eq:lagrangian}, which are essential for the generation of neutrino masses and dark matter abundance. It is clear that a complete model must contain additional fields and interactions; however, we ignore these since they are not necessarily relevant to the dark matter properties in question, introduce unnecessary model-dependence to our results, and cannot be addressed in an exhaustive manner. For instance, scenarios where the field $\phi$ is in equilibrium with the thermal bath requires significant interactions between $\phi$ and the SM particles, which might involve additional particles charged under the $U(1)'$. Likewise, in the scenarios that are supersymmetric, connecting supersymmetry breaking to the breaking of $U(1)'$ likely involves additional fields and interactions beyond the minimal ones we consider here. Such additional fields and interactions can introduce new dark matter production channels; however, such details are extremely model-dependent, hence we assume that they are subdominant to the interactions listed in Eq.\,\ref{eq:lagrangian} for the purpose of dark matter phenomenology.
Given that none of the SM particles are charged under the $U(1)'$ symmetry whereas the right-handed neutrinos are, any additional interaction connecting them must be suppressed by at least one power of $M_*$, hence such neglected interactions are expected to lead at most to $\mathcal{O}(1)$ corrections to our results. 

It is more important to consider the additional degrees of freedom that emerge from the breaking of the $U(1)'$ symmetry. If $U(1)'$ is a global symmetry that gets spontaneously broken, this introduces a light Nambu-Goldstone boson $\phi_G$, similar to the ``Majoron" from theories of spontaneously broken lepton number \cite{Chikashige:1980ui}. Its mass can be derived from an explicit soft term, which can be as high as the scale of $U(1)'$ breaking, or from quantum gravitational effects. If $\phi_G$ remains effectively massless, the same processes that lead to $N_1$ production after $U(1)'$ breaking can also lead to copious production of this light degree of freedom. If $\phi_G$ is heavier than the active neutrinos, it can decay into neutrinos before neutrino decoupling, in which case there are no observable deviations to cosmology. On the other hand, if $\phi_G$ is sufficiently long-lived, it can contribute to $N_{\text{eff}}$ as well as dark matter \cite{Baumann:2016wac,Bento:2001xi}. In our setup, the leading production mode is $\phi\rightarrow N_i\,N_i\,\phi_G$, to be compared to the dominant $N_1$ production mode $\phi\rightarrow N_1\,N_1$. From dimensional analysis, the effective vertices for these two processes are $\frac{m_\phi}{M_*}$ and $\frac{\pv}{M_*}$ respectively, whereas the former process is additionally phase space suppressed since the decay is three-body instead of two-body. Since we have $m_\phi\lsim \pv$ in this paper, the $\phi_G$ contribution to dark matter as well as $N_{\text{eff}}$ is generally subdominant to that from $N_1$.

On the other hand, the $U(1)'$, if gauged, is anomalous, and requires additional fields carrying $U(1)'$ charges in the theory (for instance, multiple copies of $\phi$) for it to be anomaly-free. Since our choice of $M_*=M_{\text{GUT}}$ is inspired by a grand unified theory at high scales, a particularly appealing UV-completion would involve fields in complete GUT multiplets, which is anomaly-free. In this case, there are additional $U(1)'$ charged fields in the theory; in particular, breaking of the $U(1)'$ introduces (massive) gauge bosons, which can play an important role in dark matter phenomenology, especially if the $U(1)'$ gauge coupling is reasonably large.

In this paper, we have assumed that such details of the underlying theory do not produce significant modifications to the dark matter properties, such that the terms in Eq.\,\ref{eq:lagrangian} capture the leading effects. 

\subsubsection{Finite Temperature Corrections}
Dark matter production takes place at high temperatures in the early Universe, where finite temperature corrections become important \cite{Drewes:2013iaa,Adulpravitchai:2015mna,Drewes:2015eoa,Garbrecht:2013gd}. Such corrections encode several relevant physical effects, such as thermal screening and Pauli blocking in the presence of a thermal plasma, effective thermal masses, and time dilation for relativistic particles; these lead to corrections of the Lagrangian level masses, vacuum decay rates, and Maxwell-Boltzmann distributions for particles that we use in our formalism. For details of how to correctly account for such effects, see \cite{Adulpravitchai:2015mna,Drewes:2015eoa}; here we simply estimate the effect of dropping these thermal corrections in our framework. 

Ref.\,\cite{Adulpravitchai:2015mna} discusses the deviations from Maxwell-Boltzmann distributions at high temperatures and subsequent effects on a population of sterile neutrinos from scalar decay. Using the results from this paper (see Appendices), we check a few representative points in our framework and find that the sterile neutrino momentum distribution and total abundance both receive $\mathcal{O}(1)$ corrections, which is consistent with the findings in Ref.\,\cite{Adulpravitchai:2015mna}. 

Beyond the distribution, individual interaction rates also receive thermal corrections for $T\gsim m_\phi$, which arise from the scalar picking up an effective thermal mass, and the decay rate receiving a time dilation correction due to the scalars being relativistic at high temperatures \cite{Adulpravitchai:2015mna,Drewes:2015eoa}. For IR dominated freeze-in scenarios, dark matter production from scalar decays is dominated by decays at $T\lsim m_\phi$, where such corrections become unimportant. Thermal corrections are more important for UV freeze-in processes for both $\phi$ and $N_1$, as they occur dominantly at the highest temperatures. UV freeze-in production of fermionic dark matter in a similar setup has been studied in \cite{Elahi:2014fsa} and \cite{McDonald:2015ljz}; the latter found that the leading thermal effect is the finite temperature quasiparticle mass $M_\phi^2=\lambda T^2/24$, where $\lambda$ is the coupling in the scalar field quartic term $\lambda \phi^4/24$, consistent with \cite{Drewes:2015eoa}. Since UV production dominantly occurs at $T\gg m_\phi$, we generally have $M_\phi\gg m_\phi$ for UV freeze-in, hence the thermal mass modification is significant. Nevertheless, $T\gg M_\phi$ still holds assuming $\lambda\,\textless\, 1$, hence we find that the scattering cross section obtained using the assumption that all particles are massless relative to the center-of-mass energy of the interaction remains a good approximation. The major thermal correction then comes from the modified phase space distribution of the initial interacting particles at high temperature; using results from \cite{Adulpravitchai:2015mna,Drewes:2015eoa}, we again estimate that these corrections are $\mathcal{O}(1)$, and can be roughly interpreted as an order of magnitude uncertainty in the value of $T_{RH}$ in our formalism; in other words, the corrections from incorporating the thermal effects in UV freeze-in can roughly be realized by shifting the value of $T_{RH}$ by an order of magnitude. We emphasize that this is an extremely rough estimate, and a proper treatment of UV freeze-in through higher dimensional operators must include a careful calculation of such thermal corrections, which currently does not exist in the literature. 

\subsubsection{Free-Streaming Length and Structure Formation}
\label{ss:fsandsf}

In this paper, we have used the free-streaming length, Eq.\,\ref{fseq}, as a measure of impact on structure formation, i.e.\,whether the dark matter candidate is cold, warm, or hot. This correspondence has recently been shown to fail for non-thermal dark matter distributions in Ref.\,\cite{Konig:2016dzg}, which studied the impact on structure formation by calculating the linear power spectrum of non-thermal dark matter distributions using the CLASS code \cite{Blas:2011rf,Lesgourgues:2011rh}, and found discrepancies between results from the two approaches. Nevertheless, we use the free-streaming length measure in this paper, so as to facilitate direct comparisons with previous literature on sterile neutrino dark matter, and because this simplistic approach is sufficient for a qualitative level of understanding of cold, warm, and hot dark matter regimes possible in the various frameworks we consider. 

As long as the dark matter distribution prominently peaks around some value and is approximately close in shape to a thermal distribution (e.g. as in Fig.\,\ref{fig:uvirps}), the concept of the ``average" velocity and free-streaming length as a representative value for the entire population remains intuitive and meaningful. On the other hand, for extremely non-thermal distributions with multiple features (such as in Fig.\,\ref{fig:s4ps}), the ``average" free-streaming length holds no meaningful information regarding structure formation; indeed, even the linear power spectrum analysis from Ref.\,\cite{Konig:2016dzg} fails to account for such extreme distributions, and the issue can only be resolved with actual numerical simulations of structure formation. In all instances where we present estimates of free-streaming length in this paper, we will use a dark matter population with a single, dominant production mode, so that the distribution peaks around some central value and does not exhibit nontrivial structures, such that the free-streaming length is still a roughly accurate reflection of the entire population. Looking at specific cases, Ref.\,\cite{Konig:2016dzg} found the free-streaming measure to be overly restrictive compared to the linear power spectrum analysis, i.e.\,the former ruled out some points as too hot when they were compatible with structure formation according to the latter analysis. Looking at the results presented in Ref.\,\cite{Konig:2016dzg}, we estimate that the discrepancy between the two approaches can be translated as roughly an order of magnitude of uncertainty in the cold/warm/hot delineations of free streaming length (for example, $\Lambda_{FS}=0.1$ Mpc could correspond to cold, warm, or hot dark matter in the linear power spectrum analysis, but $\Lambda_{FS}=1$ is most likely a hot dark matter candidate), and the free-streaming length results presented in this paper should therefore be interpreted with this degree of uncertainty in mind.

\section{Results}
\label{sec:results}

Having established our framework and formalism, in this section we present our results for each of the four scenarios of interest. For all scenarios, we assume that $N_i$ has negligible initial abundance, so that $f_{N_i} \ll1$ in the early universe; any interaction involving $N_i$ in the initial state can then be neglected, resulting in a simplification of the Boltzmann equations. The same also applies to $\phi$ abundance in scenarios where it also freezes in (Scenarios II and IV).
In scenarios where $\phi$ is in equilibrium with the SM thermal bath (Scenarios I and III), we assume that the equilibrium abundance is maintained down to some critical decoupling temperature $T_d$, below which it rapidly decays to SM radiation:
\be
\label{eq:PhiDecouple}
f_\phi(p_\phi,T) \approx
\begin{cases}
e^{-E_\phi/T} &T>T_d \\
0 & T<T_d
\end{cases}\,
\ee
We assume $T_d \approx m_\phi/20$, analogous to WIMP decoupling scenarios. Specific details of the Boltzmann equation and collision terms for each scenario are presented in Appendix \ref{sec:Collision}.

In all cases we study, we verify that the conditions for $N_1$ to freeze-in and not reach equilibrium abundance \cite{Hall:2009bx} are satisfied; for $N_1$ freezing in when $\phi$ is in equilibrium, for instance, this condition is
\beq
4\frac{M_{\text{Pl}}}{m_\phi}\left(\frac{m_{N_1}}{\pv}\right)^2\,\textless\,1.
\eeq

\subsection{Scenario I: $\phi$ in equilibrium, no supersymmetry}
\label{sec:results1}

\begin{figure}[t]
\begin{center}
\includegraphics[width=0.47\columnwidth]{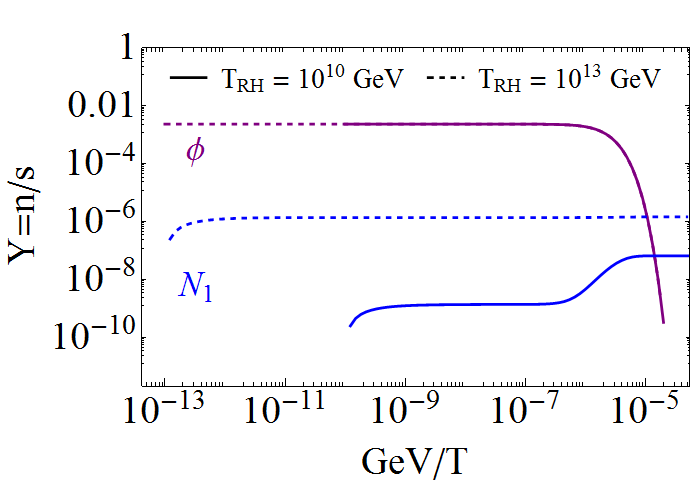}
\hskip0.3in
\includegraphics[width=0.47\columnwidth]{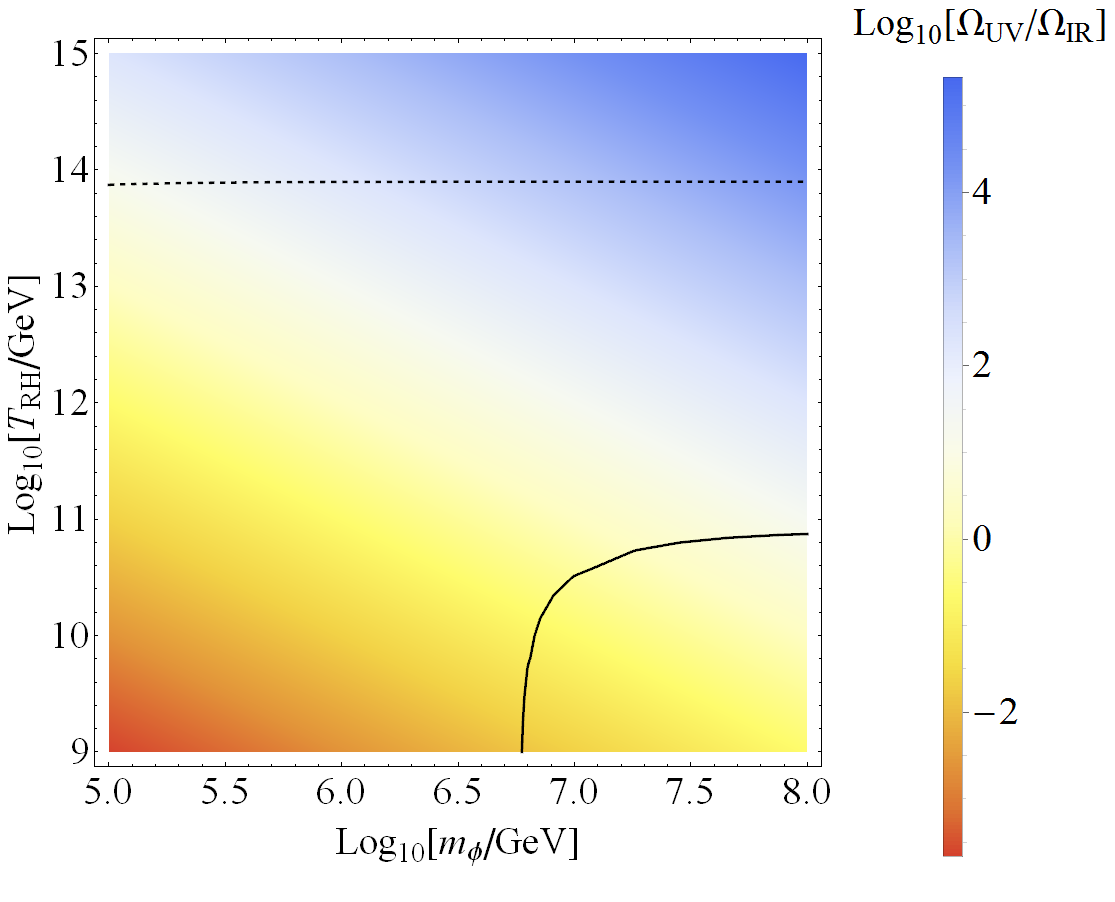}
\end{center}
\caption{Left panel: Evolution of the abundances of $\phi$ (purple) and $N_1$ (blue) for two different cases, with reheat temperatures $10^{10}$ GeV (solid lines) and $10^{13}$ GeV (dashed lines), showing IR and UV dominated production of dark matter. Here $m_\phi=1$ PeV, $\pv=100$ PeV, $m_{N_1}=1$ GeV. Right panel: The ratio of $N_1$ abundances produced from UV and IR processes. The solid and dashed lines denote where the correct dark matter abundance is achieved for $x_1 = 1,\,0.1$ ($m_N = 1, 0.1$ GeV) respectively. For this plot we set $\pv=100$ PeV. [\,Color online\,]}
\label{fig:uvir}
\end{figure}

In this scenario, the relevant processes for $N_1$ production are the UV interactions $\phi \, \phi\leftrightarrow N_1\,N_1$ and the decay process $\phi\rightarrow N_i\,N_i$ (note that the contributions from  $L_i\,H\leftrightarrow N_1\,\phi$, including all permutations, and Higgs decay $H\rightarrow L\,N_1$ are irrelevant because the corresponding Yukawa coupling $y_{i1}$ is vanishingly small). It is then interesting to see the interplay between these two contributions. Recall that the UV production rate is sensitive to the reheat temperature $T_{RH}$, and higher reheat temperatures correspond to greater $N_1$ production (see Eq.\,\ref{eq:nephidecay}). Fig.\,\ref{fig:uvir} (left panel) shows how the $\phi$ and $N_1$ abundances evolve during the early universe for two different cases, corresponding to reheat temperatures of $10^{10}$ GeV (solid curves) and $10^{13}$ GeV (dashed curves) \footnote{Plots in Fig.\,\ref{fig:uvir} are primarily intended to show the contrast between UV and IR dominated production, and do not not have the correct relic density everywhere.}. In the former case, only a small fraction of $N_1$ comes from UV freeze-in, and most of it is produced from IR freeze-in, which only turns on later, as evident from the large second bump on the solid blue curve. In the latter case with the higher reheat temperature, UV production accounts for all of the dark matter abundance, as seen in the dotted blue curve, which flattens very early. For both cases, $\phi$ tracks a thermal distribution, decouples, and then decays away (Eq.\,\ref{eq:PhiDecouple}).

In Fig.\,\ref{fig:uvir} (right panel), we show the ratio of UV to IR contributions to the final dark matter abundance for different values of the scalar mass $m_\phi$ and reheat temperature $T_{RH}$. Depending on the choice of parameters, we see that either UV or IR freeze-in can be the dominant source of $N_1$ abundance. In the UV dominated regime, the relic density should be independent of $m_\phi$ as long as $m_\phi\ll T_{RH}$, since production is dominant at higher temperatures, where $\phi$ is effectively massless. This is visible in the dotted line, which represents the contour for the correct relic density with $x_1=0.1$, and indeed does not show any $m_\phi$ dependence. On the other hand, we see that the abundance from IR production is sensitive to $m_\phi$, and decreases for larger $m_\phi$: although the decay rate grows as $\Gamma_\phi \propto m_\phi$, the time available for such decays to occur drops as $t\propto m_\phi^{-2}$, resulting in an overall decrease in abundance. This behavior is captured in the solid curve, which represents the contour for the correct relic density with $x_1=1$; as $m_\phi$ increases, this switches from being IR dominated (vertical part) to UV dominated (horizontal part).

\begin{figure}[t]
\includegraphics[width=0.6\linewidth]{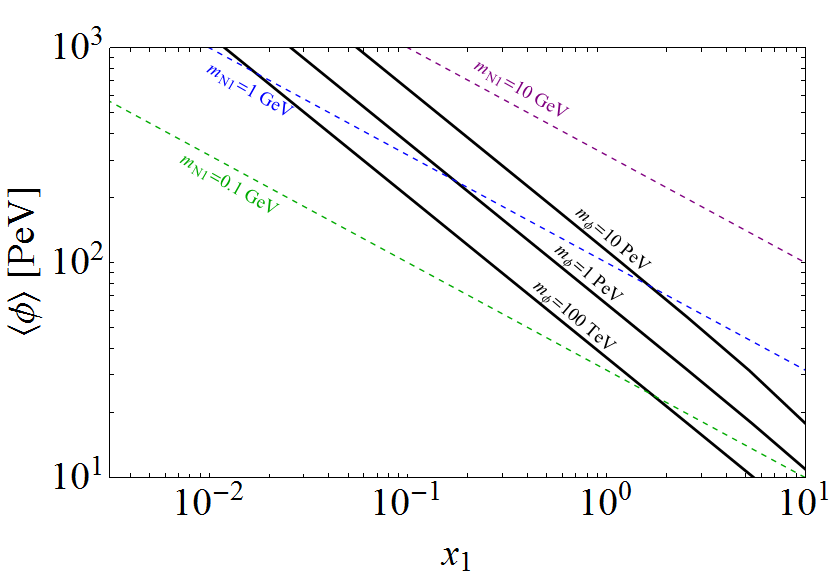}
\caption{Parameter combinations that yield the correct relic density. For each point on the plot, the correct relic density can be obtained for an appropriate choice of $m_\phi$; contours of some representative values are shown as black lines. The parameters also fix the dark matter mass; contours of various $m_{N_1}$ are shown as dotted, colored lines. Here, the reheat temperature is fixed to $T_{RH} = 10^{10}$ GeV, hence UV production is negligible.}
\label{fig:s1AbundancePlot}
\end{figure}

Next, we examine the parameter space where the correct relic abundance to account for all of the observed dark matter can be obtained. This is shown in Fig.\,\ref{fig:s1AbundancePlot} as a function of the coupling $x_1$ and the scalar vev $\pv$; for this plot, the reheat temperature is taken to be sufficiently low that only IR production is relevant. The correct relic abundance can be obtained by varying the scalar mass $m_\phi$, and the black lines show contours of various choices of $m_\phi$ for which this is achieved. For a fixed $\pv$, larger $m_\phi$ lead to lowered $N_1$ abundances, as discussed in the previous paragraph; this therefore needs to be compensated by larger couplings $x_1$, leading to a larger decay width into $N_1$ to maintain the correct abundance, as seen in the figure. These parameters also fix the mass of the dark matter particle $N_1$; in the plot, we denote contours of various $m_{N_1}$ values by colored dashed lines. This plot demonstrates that for $m_\phi$ and $\pv$ at the PeV scale, the correct DM abundance is obtained for $N_1$ at or below GeV scale masses.

\begin{figure}[t]
\includegraphics[width=0.52\linewidth]{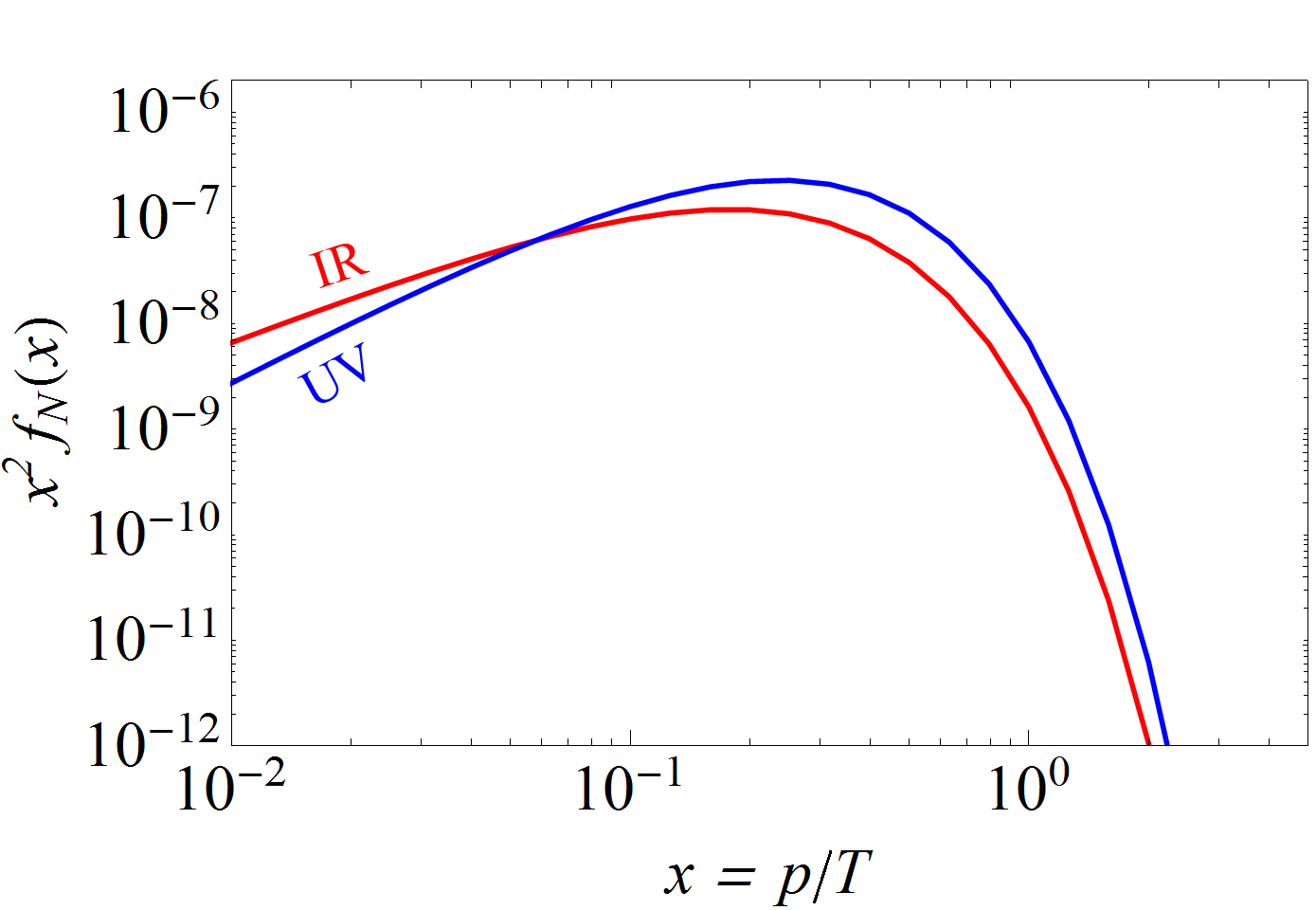}
\caption{The dark matter phase space distribution from UV (blue) and IR (red) freeze-in, for $\pv=100$ PeV, $m_\phi=1$ PeV, $m_{N_1}=1$ GeV, $T_{RH}=10^{12}$ GeV.}
\label{fig:uvirps}
\end{figure}

Next, we study the various observables related to the dark matter phase space distribution. Fig.\,\ref{fig:uvirps} shows the present distribution arising from the two production mechanisms, UV and IR freeze-in, in blue and red respectively. Despite the two production mechanisms being very different, we see from the plot that the two corresponding distributions are very similar. This similarity arises because in both mechanisms $N_1$ is produced from particles that are in equilibrium with the thermal bath, hence the characteristic energy scale at the time of production in both instances is $E_{N_1}\approx p_{N_1}\approx T$, the temperature of the bath. The UV component is slightly warmer since the annihilation rate is proportional to the center of mass energy of the process, hence dark matter is preferentially produced from interactions involving particles from the higher energy end of the equilibrium distribution. As the universe cools, the dark matter population redshifts along with the SM bath, such that $p_{N_1}\approx T$ is maintained; however, as degrees of freedom decouple, their decay products heat up the SM thermal bath but not the dark matter population, resulting in the final dark matter distribution peaking at $p_{N_1}/T\,\textless \,1$. Thus, in this scenario, the dark matter population is generally colder than the SM thermal bath. We find that this cold population results in extremely small free-streaming lengths $\Lambda_{FS}\,\textless\,10^{-4}$ Mpc and negligible contributions to $\Delta N_{\rm eff}\,\textless\,10^{-10}$.

To summarize, in this scenario, we find that dark matter can be produced with the desired relic density through a combination of UV and IR freeze-in processes, and is generally cold, so it satisfies all constraints comfortably while not showing any significant deviations from cold dark matter.

\subsection{Scenario II: $\phi$ freezes in, no supersymmetry}
\label{sec:results2}

This scenario assumes that $\phi$ does not have any significant additional interactions with the SM, and the interactions listed in Eq.\,\ref{eq:lagrangian} are therefore the ones governing its dynamics. Thus $\phi$ does not enter into equilibrium with the thermal bath in the early universe,\footnote{In such scenarios, there might exist constraints from inflationary isocurvature fluctuations, but these depend on the scale of inflation and additional self-interactions in the decoupled sector \cite{Kainulainen:2016vzv,Heikinheimo:2016yds}.} and its abundance is instead produced from freeze-in, via the UV process $L_i\,H\rightarrow N_{2,3}\,\phi$ (note that permutations of this process with $N_{2,3}$ in the initial state are absent since the heavier sterile neutrinos $N_{2,3}$ are absent in the early universe). Thus $f_{\phi} \ll 1$, and its abundance needs to be tracked using the Boltzmann equations. This frozen-in population of $\phi$ then decays entirely into sterile neutrinos once $\phi$ obtains a vev, as there are no competing decays into SM particles, thereby producing dark matter via $\phi\rightarrow N_1\,N_1$. Note that the UV freeze-in process $\phi\phi\rightarrow N_1N_1$ is inactive here due to the suppressed abundance of $\phi$ at high temperatures. Details of the Boltzmann equations and collision terms are again presented in Appendix \ref{sec:Collision}.

\begin{figure}[t]
\includegraphics[width=0.47\columnwidth]{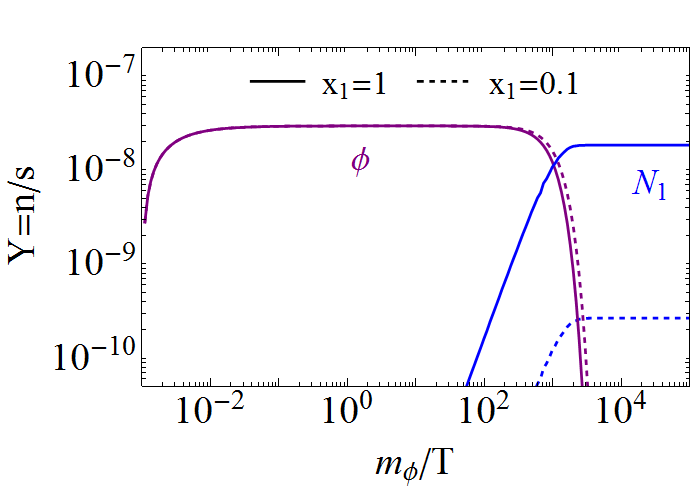}
\hskip0.3in
\includegraphics[width=0.47\columnwidth]{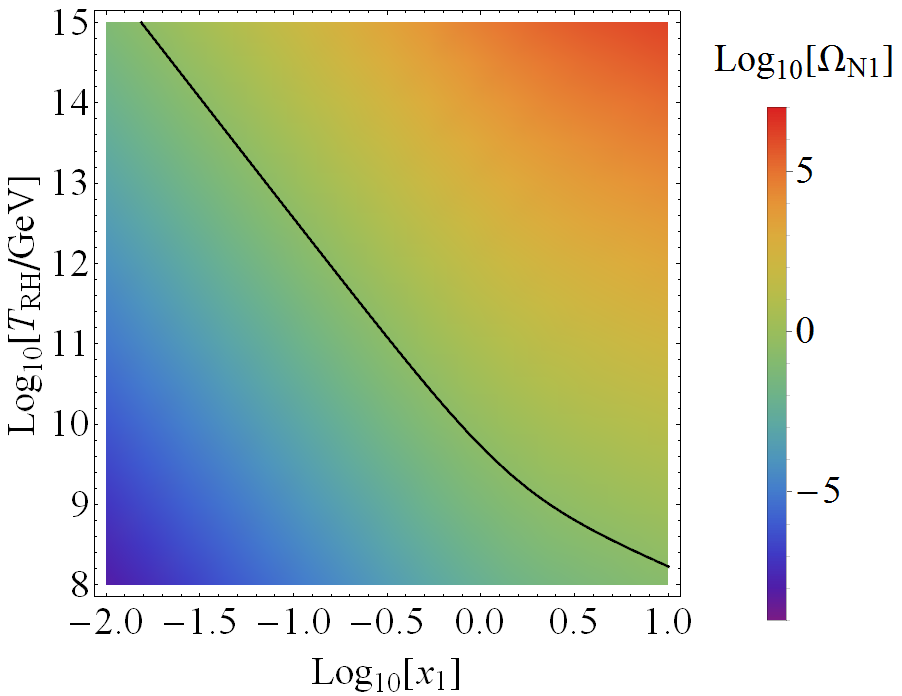}
\caption{Left panel: Evolution of the yields of $\phi$ and $N_1$ in the early universe, for some fixed $m_\phi$. Here we have fixed $T_{RH}= 10^{10}$ GeV and $\pv=100$ PeV, and show the evolution for two different values of $x_1$. Right panel: The dark matter relic abundance $\Omega_{N_1}$ as a function of the coupling $x_1$ and the reheat temperature $T_{RH}$; the black curve denotes the combinations that result in the correct relic abundance $\Omega_{N_1} h^2=0.12$. Here, we have set $\pv=100$ PeV. [\,Color online\,]}
\label{fig:s2yieldrelic}
\end{figure}

This freeze-in of $\phi$ and subsequent decay to $N_1$ is illustrated in the left panel of Fig.\,\ref{fig:s2yieldrelic}. We illustrate this process for two different choices of the coupling $x_1$, which controls the branching fraction into $\phi\rightarrow N_1\,N_1$ and therefore the final dark matter abundance. The plot shows two distinct features as the coupling gets larger: (i) a larger abundance of $N_1$, consistent with Br($\phi\rightarrow N_1\,N_1)\propto x_1^2$, and (ii) a more rapid depletion of $\phi$, since a larger $x_1$ also results in a larger $\phi$ decay width. Thus, the final dark matter abundance is set by the freeze-in abundance of $\phi$, which depends on $T_{RH}$, and the branching fraction $\phi \rightarrow N_1\,N_1$, which depends on the $x_1$ coupling, with a larger value of either parameter resulting in a larger abundance \footnote{Note that neither the UV freeze-in of $\phi$ nor the branching fraction into $N_1$ is sensitive to $m_\phi$ as long as $m_\phi\gg m_{N_i}$, hence the exact value of this parameter is irrelevant.}. This behavior is illustrated in the right panel of Fig.\,\ref{fig:s2yieldrelic}, which shows how the dark matter relic abundance depends on the values of these parameters. The black curve denotes the combinations that result in the correct relic abundance $\Omega_{N_1} h^2=0.12$; the curve changes slope around $x_1=1$ as $\phi$ switches from decaying dominantly into $N_{2,3}$ at lower values of $x_1$ to decaying primarily into $N_1$ at higher values. Thus, we see that even when both $\phi$ and $N_1$ are absent in the early universe, the desired dark matter abundance can be built up with a sufficiently high reheat temperature to produce $\phi$ from freeze-in and an appropriate coupling $x_1$ to convert a fraction of the $\phi$ population into $N_1$.

\begin{figure}[t]
\includegraphics[width=0.48\columnwidth]{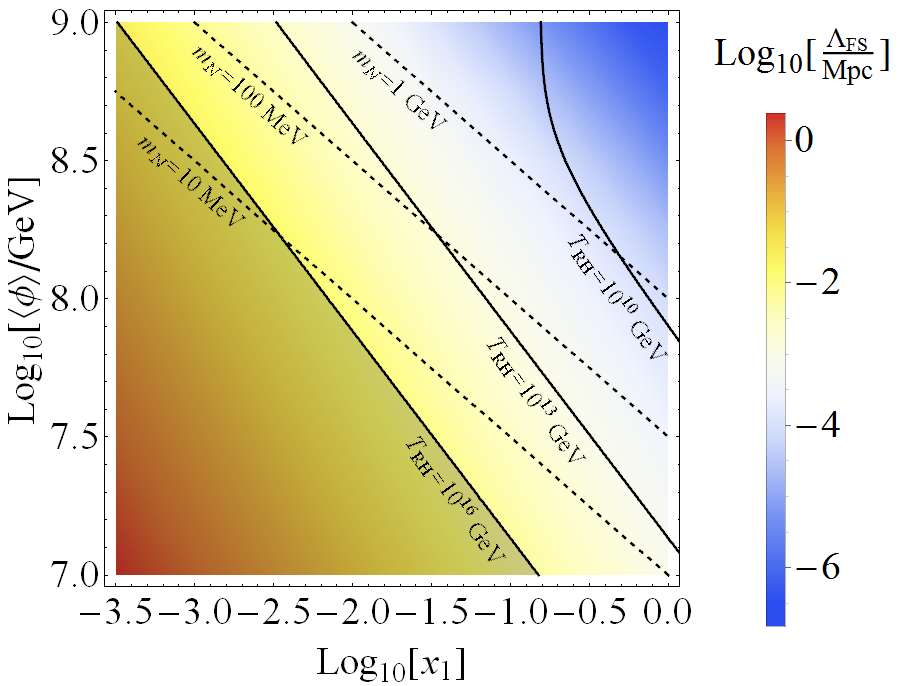}
\hskip0.1in
\includegraphics[width=0.49\columnwidth]{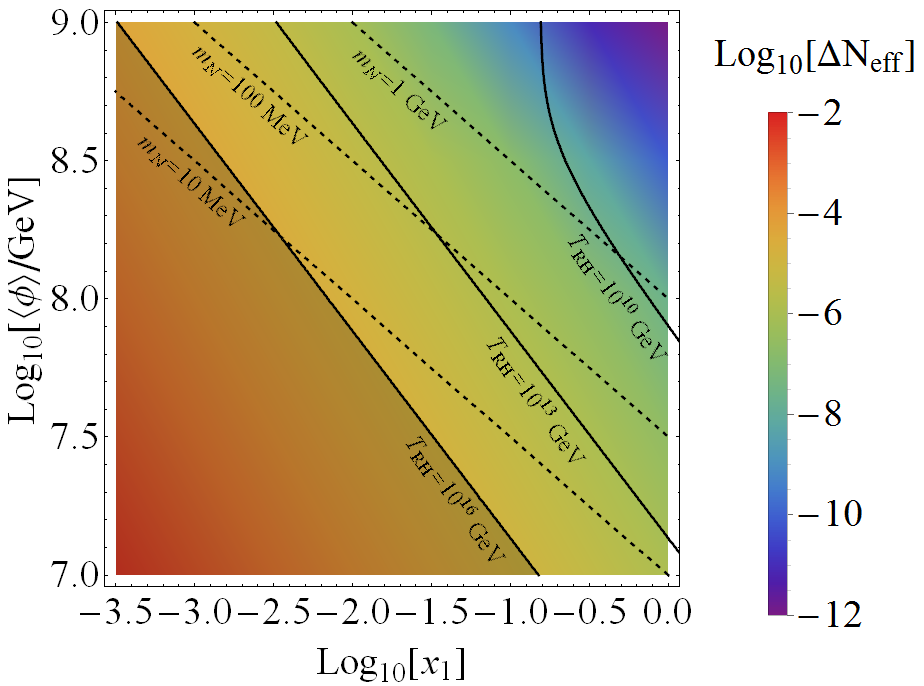}
\caption{The free-streaming length $\Lambda_{FS}$ (left panel) and the contribution to the effective number of relativistic degrees of freedom at BBN (right panel) [\,Color online\,]. The dashed lines show contours of $m_{N_1}$. In these plots, the correct relic density is achieved by appropriately choosing $T_{RH}$; some contours of the required $T_{RH}$ are shown as solid curves. The shaded regions are not accessible since the required $T_{RH}$ here is greater than the GUT scale, requiring the theory studied here to be UV completed.}
\label{fig:FSNeffPlot}
\end{figure}

In this scenario, $\phi$ is fairly long-lived since its decay width is suppressed due to the small effective couplings $\sim 2 x_i\pv/M_*$ to the sterile neutrinos. Thus, its decay produces $N_1$ particles with energies of order $m_\phi$ at late times, when the temperature of the ambient bath is significantly lower. This behavior is already visible in the left panel of Fig.\,\ref{fig:s2yieldrelic}, where we see that $E_{N_1}\sim p_{N_1}\sim m_\phi \gg T$ at the time of production ($\ie$ where the $\phi$ yield drops). The $\phi$ lifetime can be extended by suppressing these effective couplings, which can be accomplished by lowering either $\pv$ or $x_i$, which results in warmer dark matter.  In Fig.\,\ref{fig:FSNeffPlot}, we plot the free-streaming length $\Lambda_{FS}$ and the contribution to the effective number of relativistic degrees of freedom $\Delta N_{\rm{eff}}$(BBN) for these parameters. On both plots, we set the relic density to the correct value by appropriately choosing $T_{RH}$; some contours of the required $T_{RH}$ values are shown on the plots as solid lines. Both plots show that dark matter becomes hotter as these parameters are lowered; however, in the shaded region, the correct relic density cannot be achieved without reheating above the GUT scale, where our theory needs to be UV completed, hence the ``hot" regions in the bottom left corners of the plots are not accessible. In the allowed region, we see that it is possible for dark matter produced in this scenario to be warm, and $\Delta N_{\rm{eff}}\lsim 10^{-4}$.

\subsection{Scenario III: $\phi$ in equilibrium, supersymmetry}
This supersymmetric extension of Scenario I introduces new particles and interactions that can contribute to the production of $N_1$. Here, we assume that $\psi$ (the fermionic superpartner of $\phi$) is in equilibrium in the early universe via the supersymmetric counterparts of the interactions that keep $\phi$ in equilibrium, and decay away rapidly once out of equilibrium. Overall, the processes that contribute to dark matter production in this scenario are
\begin{align}
\text{UV:}& ~~ \phi\,\phi \rightarrow N_1\, N_1\,,~~~ \phi \,\psi \rightarrow \tilde{N}_1\, N_1 \,,~~ ~\psi \,\psi \rightarrow \tilde{N}_1 \, \tilde{N}_1\,  \nonumber\\
\text{IR:}& ~~ \phi \rightarrow N_1 \, N_1\,,~~~\phi\rightarrow \tilde{N}_1\,\tilde{N}_1,~~~ \tilde{N}_1 \rightarrow \psi \, N_1\, .
\end{align}
Note that we do not consider the UV process $\phi\,\phi\rightarrow \tilde{N}_1\,\tilde{N}_1$ that arises from the soft term proportional to $A_\eta$ from Eq.\,\ref{eq:softterms} as it only turns on at relatively low temperatures (after supersymmetry is broken), whereas we do consider its IR counterpart $\phi\rightarrow \tilde{N}_1\,\tilde{N}_1$, which can be important if $A_\eta$ is comparable to or larger than $m_\phi$. The relevant Boltzmann equations and collision terms are presented in Appendix \ref{sec:Collision}.

\begin{figure}[t]
\includegraphics[width=0.5\linewidth]{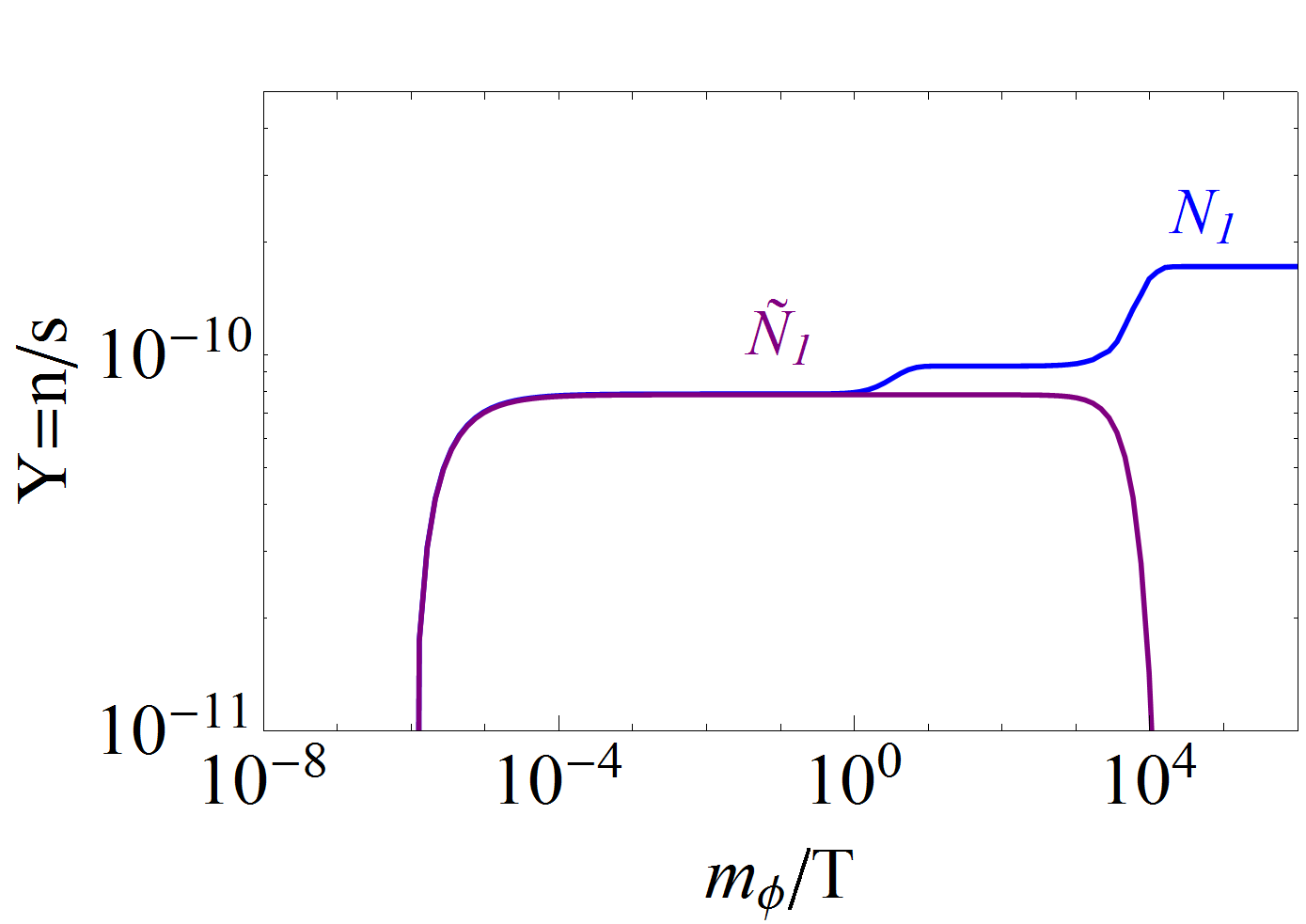}
\caption{The yields of $N_1$ (blue) and $\tilde{N}_1$ (purple) during freeze-in. Three distinct dark matter production phases are visible: an early UV freeze-in of both $N_1$ and $\tilde{N}_1$, $\phi$ decay, and $\tilde{N}_1$ decay. Here we have set $T_{RH} = 10^{12}$ GeV, $m_\phi = 1$ PeV, $m_{\tilde{N}_1} = 16$ PeV, $\pv=100$ PeV, and $m_{N_1} = 40$ MeV. For this plot we have assumed that $A_{\eta_1}$ is negligible, so that there is no appreciable production of $\tilde{N}_1$ from $\phi\rightarrow\tilde{N}_1\tilde{N}_1$.}
\label{fig:s3YieldPlot}
\end{figure}

In this scenario, $\phi$ and $\psi$ are in equilibrium, whereas $N_1$ and $\tilde{N}_1$ freeze-in. $\phi$ and $\tilde{N}_1$ both decay (in and out of equilibrium respectively), leading to a period of IR freeze-in for $N_1$. This process is illustrated in Fig.\,\ref{fig:s3YieldPlot}, where we plot the evolution of the yields of $N_1$ and $\Nt$. Note that three distinct phases of $N_1$ production are clearly visible in the plot. An early UV freeze-in phase occurs at $m_\phi/T \leq 10^{-4}$; here, the $N_1$ and $\tilde{N}_1$ production mechanisms are identical, hence their abundances trace the same curve. Next, a second bump in $N_1$ abundance occurs around $m_\phi/T \sim 1$ from $\phi$ decay. Finally, there is another bump corresponding to contributions from $\tilde{N}_1$ decay at late times, around $m_\phi/T \sim 10^4$, reflecting the relatively long lifetime of $\tilde{N}_1$. Depending on the choice of parameters, these three different production mechanisms can contribute different amounts of dark matter. UV production is dominant when $T_{RH}$ is large; in this case, equal amounts of $N_1$ and $\tilde{N}_1$ are produced, resulting in dark matter made up equally of $N_1$ from UV freeze-in and $\tilde{N}_1$ decay. If $T_{RH}$ is low, IR production is dominant; in this case, $N_1$ can be produced directly from $\phi$ decay or from the decay of $\tilde{N}_1$ produced via $\phi\rightarrow\tilde{N}_1\tilde{N}_1$. For these two decay widths, $\Gamma(\phi\rightarrow N_1 N_1)\propto \eta_{1eff}^2 m_\phi$ and $\Gamma(\phi\rightarrow\tilde{N}_1\tilde{N}_1)\propto \eta_{1eff}^2 A_{ \eta_1}^2/m_\phi$, hence the former (latter) contribution dominates for $m_\phi\,\textgreater \,A_{\eta_1}$ ($m_\phi\,\textless \,A_{\eta_1}$). In the latter case, it is therefore possible for the entire dark matter abundance to originate from $\tilde{N}_1$ decay.

\begin{figure}[t]
\includegraphics[width=0.52\linewidth]{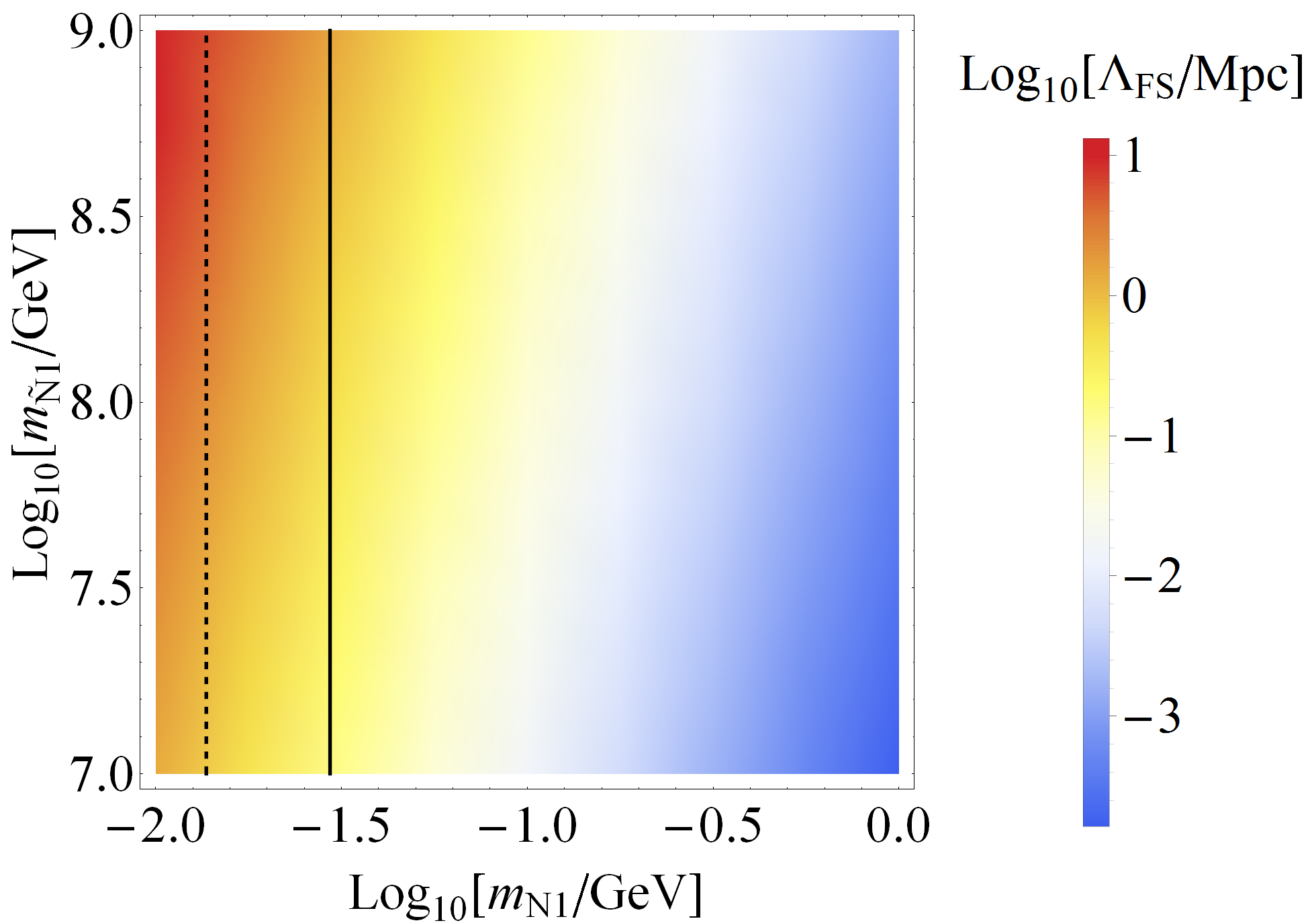}
\caption{Free-streaming length $\Lambda_{FS}$  [\,Color online\,]. The solid (dotted) line denotes where $N_1$ makes up $100\%$ ($10\%$) of the total dark matter abundance (for $\pv=100$ PeV, $T_{RH}=10^{15}$ GeV). Cold, warm, and hot dark matter are all viable options in this scenario.}
\label{fig:s3FSCountourPlot}
\end{figure}

While the free-streaming length and $\Delta N_{\rm eff}$ contribution from UV production and $\phi$ decay follow the same patterns as in Scenario I, the presence of a new production channel in the form of $\tilde{N}_1$ decay opens additional possibilities. Because $\tilde{N}_1\rightarrow \psi N_1$ is the $\emph{only}$ available decay channel, suppressing the corresponding coupling can make $\tilde{N}_1$ extremely long-lived and the subsequently produced $N_1$ extremely hot (note that this is not possible with $\phi$, since its lifetime is determined by other decay channels such as $\phi\rightarrow N_{2,3} N_{2,3}$). To illustrate this, we plot the free-streaming length as a function of $N_1$ and $\tilde{N}_1$ masses in Fig.\,\ref{fig:s3FSCountourPlot}. The solid (dotted) line denotes combinations resulting in $N_1$ making up $100\%(10\%)$ of the total dark matter abundance (for $\pv=100$ PeV, $T_{RH}=10^{15}$ GeV). The figure shows that the parameter space allows for hot (inconsistent with structure formation), warm, or cold dark matter. Constraining $\Lambda_{FS} \lesssim 0.1$ Mpc, we find $\Delta N_{\rm eff} \lesssim 10^{-4}$ if $N_1$ comprises all of dark matter; this is consistent with Scenario II above and with results in Ref.\,\cite{Merle:2015oja}, which found that large $\Delta N_{\rm eff}$ during BBN is inconsistent with free-streaming length constraints.

An interesting possibility worth entertaining is the case where late decays of $\tilde{N}_1$ result in only a tiny fraction ($\textless\, 1\%$; see $\eg$ Ref.\,\cite{Hooper:2011aj}) of (extremely hot) dark matter, while the rest of the dark matter (either Higgsino or $N_1$ from $\phi$ decay) is cold. In this case, this subdominant population of $N_1$ from $\tilde{N}_1$ decays is not subject to any free-streaming constraints (since the bulk of dark matter is cold), but can still provide a large contribution to $\Delta N_{\rm eff}$ if $\tilde{N}_1$ is sufficiently heavy and long-lived (but decays before LSP decoupling). We find that these conditions are satisfied for a heavy $\tilde{N}_1$ and an extremely light $N_1$. However, a heavy $\Nt$ requires an even heavier $\phi$ (if $\tilde{N}_1$ is to be produced via $\phi\rightarrow\tilde{N}_1\,\tilde{N}_1$), which does not allow enough time for sufficient IR freeze-in of $\tilde{N}_1$, as this process ends once $\phi$ goes out of equilibrium. Alternatively, one can consider dominantly UV production of $\tilde{N}_1$ via $\psi\psi\rightarrow\tilde{N}_1\tilde{N}_1$; however, this goes through the coupling $\eta_1$, which is proportional to $m_{N_1}$, hence raising $\eta_1$ to increase $\tilde{N}_1$ production also raises $m_{N_1}$, reducing $\Delta N_{\rm eff}$. Therefore, while this idea is in principle feasible, we find that the relations between various parameters imposed by our framework do not allow us to fully realize this attractive possibility, and we obtain at most $\Delta N_{\rm eff}\sim 10^{-3}$ in this scenario in our framework. However, we note that such observationally interesting $\mathcal{O}(0.1)$ values of $\Delta N_{\rm eff}$ at BBN may indeed be realized in a more general framework \cite{Shakya:2016oxf}.

\subsection{Scenario IV: $\phi$ freezes in, supersymmetry}

In this section, we will assume that the heavier sterile neutrinos $N_2, N_3$ are sufficiently heavy that the entropy dilution from their decay is negligible. This scenario is a supersymmetric extension of Scenario II, and therefore shares many of the features from Scenarios II and III above. For the freeze-in of $\phi$, compared to Scenario II we have the following additional interactions:
\begin{center}
\begin{tabular}{c c c}
$\tilde{L}_i\,H\rightarrow \tilde{N}_{2,3}\,\phi$\,,~~ & $L_i\,\tilde{H}\rightarrow \tilde{N}_{2,3}\,\phi$\,,~~ & $\tilde{L}_i\,\tilde{H}\rightarrow N_{2,3}\,\phi$\,,
\end{tabular}
\end{center}
since the charged and neutral Higgsinos and sleptons are also present in the thermal bath. Similar processes also lead to UV production of $\psi$, which subsequently decay as $\psi \rightarrow \tilde{L}_i\, H \, N_{2,3}$ or $\psi \rightarrow \tilde{N}_{2,3}\, N_{2,3}$. Again, one must ensure that the decays of all supersymmetric particles occur before Higgsino decoupling. As $\phi$ and $\psi$ are absent in the early universe, there is no direct UV production of $N_1$ or $\tilde{N}_1$, and dark matter is produced via the decay processes
\beq
\phi \rightarrow N_1 \, N_1\,;~~~\phi\rightarrow \tilde{N}_1\,\tilde{N}_1,~~~ \tilde{N}_1 \rightarrow \psi \, N_1\, .
\eeq

The full set of Boltzmann equations and collision terms are presented in Appendix \ref{sec:Collision}.

\begin{figure}[t]
\includegraphics[width=0.52\linewidth]{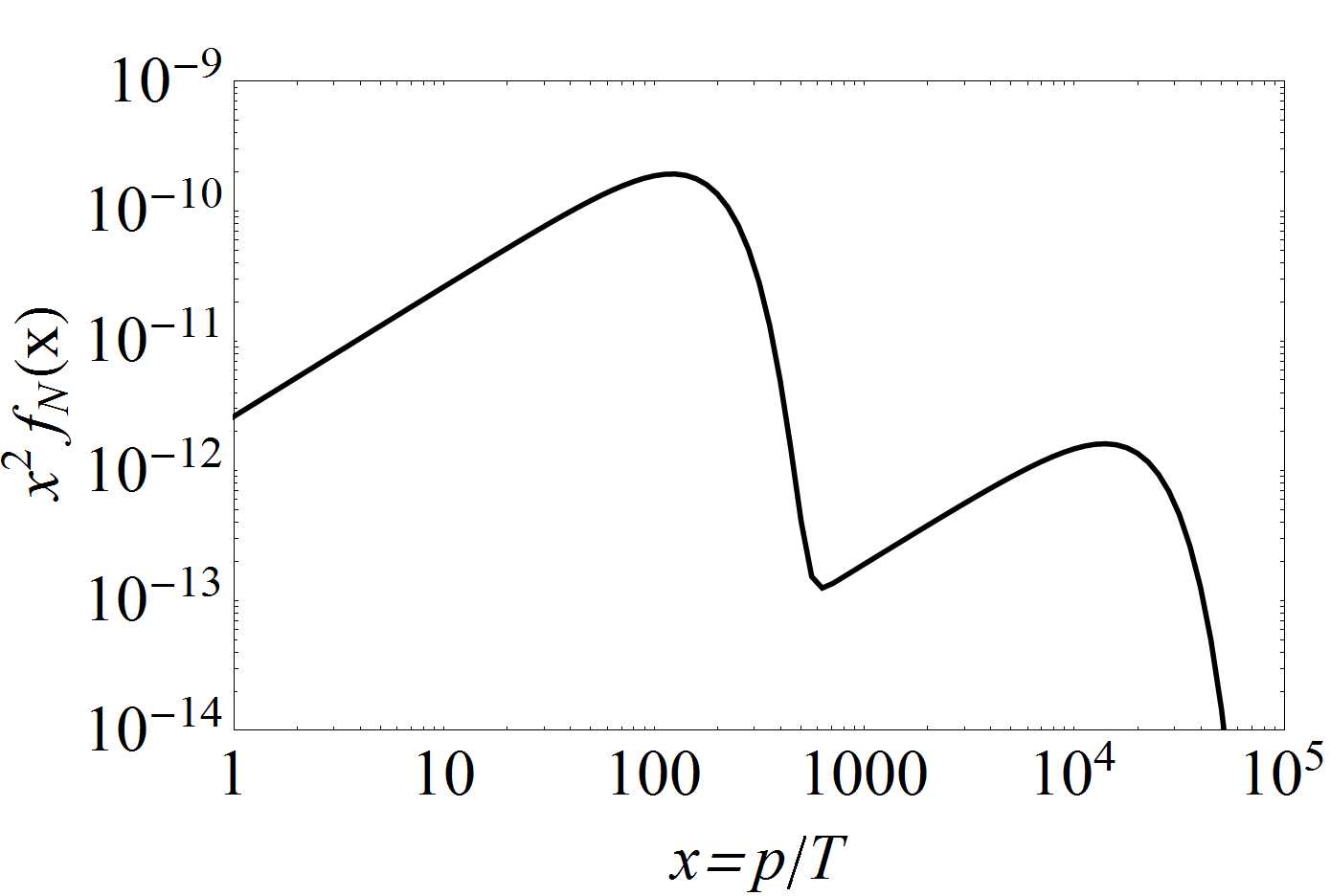}
\caption{Phase space distribution for a case with comparable scalar and sterile sneutrino decay contributions. In this plot, the parameters are: $m_N = 1$ GeV, $m_{\tilde{N}_1} = 10^8$ GeV, $m_\phi = 10^9$ GeV, $m_\psi = 10^7$ GeV, $A_{\eta1} = 10^9$ GeV, $\pv = 10^9$ GeV.}
\label{fig:s4ps}
\end{figure}

Here, $\phi$ does not have any other significant interactions and therefore decays primarily to $N_i$ and $\tilde{N}_i$, while the presence of $N_i$ allows for late decays into extremely energetic $N_1$. The phase space distribution of $N_1$ produced in this manner is shown in Fig.\,\ref{fig:s4ps}, with the parameter choices as described in the plot caption. We see that there are two distinct bumps in this particular distribution: the lower momentum one corresponds to $N_1$ produced directly from $\phi$ decays, while the higher momentum bump corresponds to the contribution from $\tilde{N}_1$ decays. The two bumps peak at $x\sim 100$ and $x\sim 10^4$, reflecting that both arise from late decays where the mass of the decaying particle is several orders of magnitude higher than the temperature of the ambient thermal bath. In such scenarios, we therefore see that we can get extremely nontrivial phase space distributions of warm/hot dark matter, which might prove to be of interest for various considerations.

\begin{figure}[t]
\includegraphics[width=0.6\linewidth]{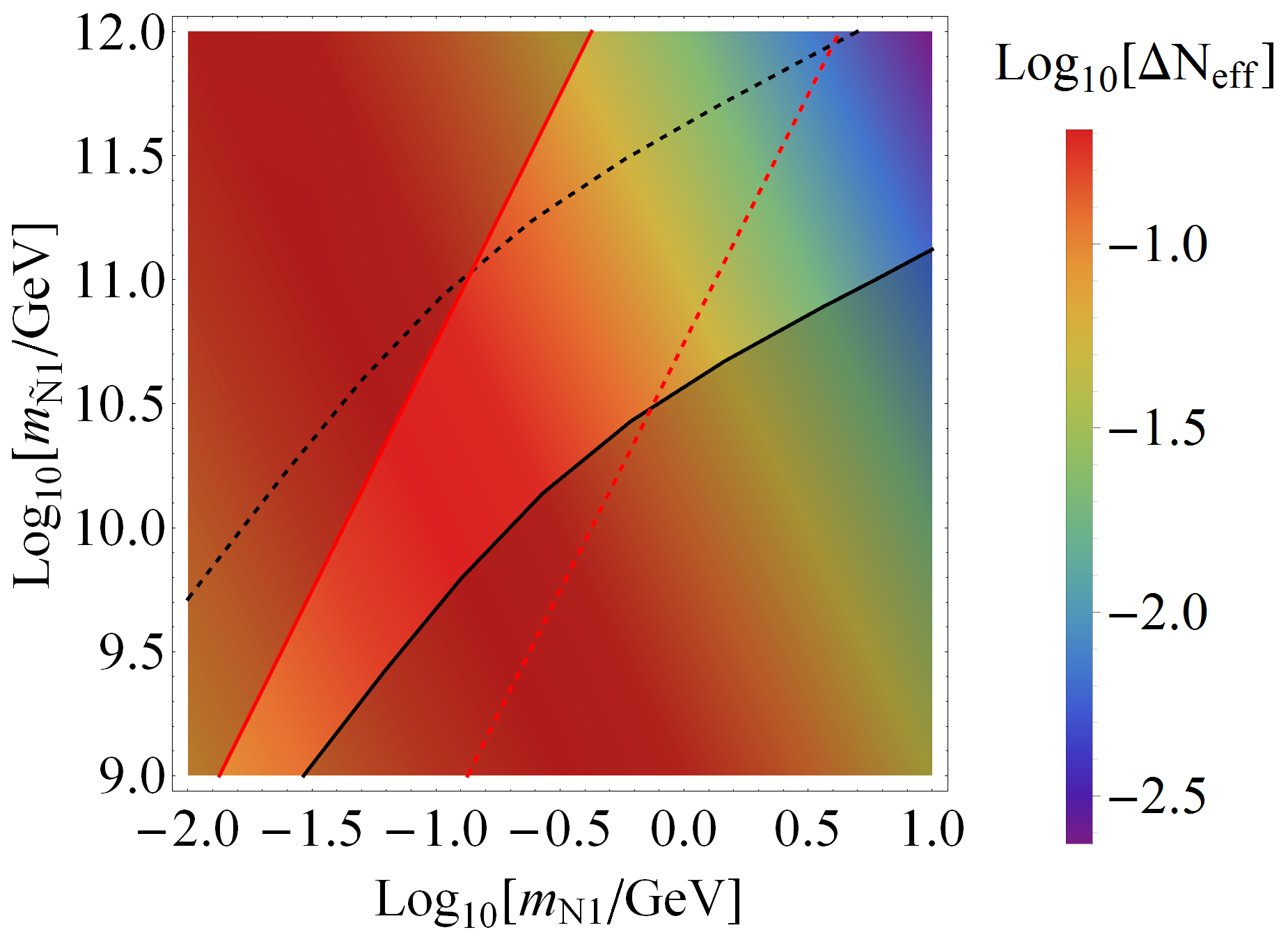}
\caption{Relic density and $\Delta N_{\rm eff}$ (BBN) for $T_{RH}=10^{15}$ GeV and $\pv=0.1 m_{\tilde{N}_1}$. In this plot, $\phi$ decays dominantly to $\tilde{N}_1$, and the decays of $\tilde{N}_1$ populate $N_1$. Solid (dashed) black curves denote where $N_1$ accounts for $100\% (1\%)$ of dark matter; the solid (dashed) red lines denote where the decay occurs at the decoupling temperature of a Higgsino of mass 200 (2000) GeV. Shaded regions are disallowed because of overclosure (bottom right) or $N_1$ decaying after a 200 GeV Higgsino freezes out (top left). [\,Color online\,]}
\label{fig:s4NeffPlot}
\end{figure}

As in the previous scenarios, the correct relic density can be obtained with appropriate choices of the various parameters, combining the multiple production mechanisms for dark matter; since the patterns are mostly the same as in Scenarios II and III, we do not repeat those details again. Given the energetic nature of the dark matter particles produced from out of equilibrium decays, it is more interesting to study the observational properties of such a population. As in Scenario III, cold, warm, and hot dark matter are all possible in this scenario. In addition, we find that contributions to $\Delta N_{\rm eff}$ at BBN with a subdominant ($1\%$) fraction of dark matter, as discussed in the final paragraph in Scenario III, has better prospects in this scenario as $\phi$ can decay to $N_1$ and $\tilde{N}_1$ out of equilibrium. For a proof of concept, we focus on the case where $A_{\eta_1}\gg A_{\eta_{2,3}},\,m_\phi$, so that the entire population of $\phi$ that freezes in decays into $\tilde{N}_1$. In this case, the entire population of $N_1$ is produced from $\tilde{N}_1$ decays. The remainder (dominant fraction) of dark matter should then be accounted for by some other component, $\eg$ the Higgsino. We plot the $\Delta N_{\rm eff}$ and relic density obtained with these approximations in Fig.\,\ref{fig:s4NeffPlot}. The color coding denotes the size of $\Delta N_{\rm eff}$; the black curves and red lines denote contours of relic density and decoupling temperature respectively, as explained in the caption. Shaded regions are disallowed because of overclosure (bottom right) or $N_1$ decaying after a 200 GeV Higgsino freezes out (top left). In the allowed (non-shaded) region, even imposing that $N_1$ make up less than $1\%$ of dark matter ($\ie$ region above the dashed black curve), we see that it is possible to get  $\Delta N_{\rm eff}\approx 0.1$, which is an extremely interesting feature that can potentially be probed by future measurements.

\section{Summary}
\label{sec:summary}

In this paper, we have investigated cosmological aspects of light ( $\lsim$ GeV scale) sterile neutrino dark matter produced from the freeze-in mechanism. Given that such a dark matter candidate interacts feebly with the SM and thus has no promising indirect or direct search strategies, such cosmological aspects represent the most phenomenologically interesting features of such a candidate. While previous papers have performed similar studies in more restricted setups, we perform this study in a comprehensive framework that includes many interesting variations: production from a scalar in or out of equilibrium with the thermal bath in the early universe, via UV or IR freeze-in, and with or without supersymmetry. Under this broad approach, we find many novel features that were missed by earlier studies. Our findings can be summarized as follows:

\begin{itemize}
\item Relic density: The relic abundance required to explain all of dark matter can be achieved in all scenarios considered. Production can occur dominantly through UV freeze-in, IR freeze-in from decays of the scalar $\phi$ in or out of equilibrium with the SM bath, or through decays of a sterile sneutrino in supersymmetric setups; more generally, any combination of these processes can also result in the observed relic density.

\item Free-streaming length: We find that sterile neutrino dark matter produced via freeze-in can be cold, warm, or hot, depending on the dominant production mechanism and choice of parameters. Dark matter from UV production or decay of $\phi$ in equilibrium with the thermal bath is generally cold (Scenario I), while late out of equilibrium decay of $\phi$ or the sterile sneutrino $\tilde{N}_1$ can result in warm or hot dark matter (Scenarios II, III, IV). Such scenarios can be of great interest from the point of view of structure formation.

\item Phase space distribution: Given the interplay of multiple production mechanisms for dark matter, its momentum distribution can be extremely varied and nontrivial. UV and IR freeze-in produce dark matter with slightly different momentum distributions (Fig.\,\ref{fig:uvirps}); likewise, dark matter produced from decays of $\phi$ (in or out of equilibrium) and $\tilde{N}_1$ can have significantly different distributions if the times and energy scales of decay are very different (see Fig.\,\ref{fig:s4ps}). Note that such distributions are possible only because the $N_1$ abundance freezes in and only has feeble SM and self interactions, hence different components produced from different mechanisms do not mix but maintain their individual phase space distributions. Such features are not present in the traditionally studied dark matter candidates that freeze out of equilibrium.

\item Contributions to $\Delta N_{\rm eff}$ during BBN: Extremely energetic dark matter particles in the early universe can mimic dark radiation, contributing to the effective number of relativistic degrees of freedom $\Delta N_{\rm eff}$. For GeV scale sterile neutrinos, we find that such contributions are more likely at BBN than CMB since they redshift and become non-relativistic at later times. We find that $\Delta N_{\rm eff}$ is generally restricted to negligible values ($\lsim 10^{-4}$) by free-streaming length constraints if $N_1$ makes up all of dark matter ($\eg$ Fig.\,\ref{fig:FSNeffPlot}). However, free-streaming constraints can be circumvented if $N_1$ makes up only a subdominant fraction (\,$\lsim 1\%$\,) of dark matter, and in this case we find that $\Delta N_{\rm eff}\sim\mathcal{O}(0.1)$ can indeed be realized consistent with all other constraints (see Fig.\,\ref{fig:s4NeffPlot}).

\end{itemize}

Finally, while we performed the above study in a specific framework, so that many of the quantitative results are model-dependent, we emphasize that the general features discussed here represent the most observable aspects of frozen in sterile neutrinos, and are more broadly applicable to any framework that has such a candidate.

\medskip
\textit{Acknowledgements: }We thank James D. Wells for collaboration in the early stages of the project, and for valuable discussions and suggestions on the manuscript. The authors are supported in part by the DoE under grants DE-SC0007859 and DE-SC0011719. This work was performed in part at the Aspen Center for Physics, which is supported by National Science Foundation grant PHY-1066293.

\bibliography{neutrinocosmologybib}

\appendix
\section{Boltzmann Equations and Collision Terms for Various Scenarios}
\label{sec:Collision}

\subsection{Scenario I: $\phi$ in equilibrium, no supersymmetry}

Assuming that $N_i$ has a negligible initial abundance, the relevant phase space density weights in the Boltzmann equations simplify to
\beq
\Omega(\phi\rightarrow N_1\,N_1) \approx f_\phi,~~~\Omega(\phi\,\phi\rightarrow N_1\,N_1) \approx f_\phi f_\phi\,,
\eeq
resulting in the following Boltzmann equation for $f_{N_1}$:
\be
H r \frac{\partial f_{N_1}}{\partial r} = C_{\phi\rightarrow N_1\,N_1}[f_{N_1}] + C_{\phi\,\phi\rightarrow N_1\,N_1}[f_{N_1}]\,.
\ee

Each collision term takes the form given in Eq.\,\ref{eq:collision}. Since both process take place for temperatures much greater than the mass $m_{N_1}$, we will approximate $N_1$ as massless throughout the calculation. Furthermore, since the annihilation of $\phi$ is a UV process taking place only a high temperatures just after reheating, we will set $m_\phi\rightarrow 0$ for the computation of the annihilation collision term.

The collision term for the annihilation process is
\begin{align}
C_{\phi\,\phi\rightarrow N_1\,N_1}[f_{N_1}] &= \frac{2}{2E_{N_1}} \int \diff \Pi_\phi \diff\Pi_{\phi'} \diff \Pi_{N_1'} \M (2\pi)^4 \delta^4\left(\Sigma p\right) f_\phi(p_\phi)  f_\phi(p_{\phi'}) \\
&=\frac{x_1^2}{(2\pi)^3 M_*^2} ~T^3 \exp(-\pn/T) ~\Theta(T-T_d)\,. \label{eq:s1annihilation}
\end{align}
where $T_d$ is the decoupling temperature of $\phi$ (Eq.\,\ref{eq:PhiDecouple}).

The collision term corresponding to $\phi$ decay is
\begin{align}
C_{\phi\rightarrow N_1\,N_1}[f_{N_1}] &= \frac{2}{2E_{N_1}} \int \diff \Pi_\phi \diff \Pi_{N_1'} \M (2\pi)^4 \delta^4\left(\Sigma p\right) f_\phi(p_\phi) \\
&= \frac{ \xeff^2 m_\phi^2}{16 \pi\pn^2} \int_{p_{\phi,\text{min}}}^\infty \diff p_\phi \frac{p_\phi}{E_\phi} f_\phi(p_\phi)\,,  \label{eq:phidecaysimplified}
\end{align}
where $\xeff = \frac{2 x_1 \pv}{M_*}$ and kinematic considerations restrict the momentum integration over $p_\phi$ to be greater than
\be
p_{\phi,\text{min}} \equiv \left| \pn - \frac{m_\phi}{4\pn} \right| \,.
\ee

\subsection{Scenario II: $\phi$ freezes in, no supersymmetry}

Since $\phi$ freezes in, $f_{\phi} \ll 1$, and for its UV production process $L_i\,H\rightarrow N_{2,3}\,\phi$ we can approximate
\beq
\Omega(L_i\,H\rightarrow N_{2,3}\,\phi) \approx f_H f_{L_i} \approx e^{-(E_{L_i}+E_H)/T}\,.
\eeq

This frozen-in population of $\phi$ decays entirely into sterile neutrinos once $\phi$ obtains a vev. Such decays lead to DM production via $\phi\rightarrow N_1\,N_1$, with
\beq
\Omega(\phi\rightarrow N_1\,N_1) \simeq f_\phi\,.
\eeq
The corresponding Boltzmann equations for the freeze-in of $\phi$ and subsequent production of dark matter $N_1$ are
\begin{align}
H r \frac{\partial f_\phi}{\partial r} &= \sum_{\substack{i=e,\mu,\tau\\ j=2,3}} C_{L_i\,H\rightarrow\, N_j\,\phi}[f_\phi] + \sum_{k=1,2,3} C_{\phi\rightarrow N_k\,N_k}[f_\phi] \label{eq:BoltzPhi}\\
H r \frac{\partial f_{N_1}}{\partial r} &= C_{\phi\rightarrow N_1\,N_1}[f_{N_1}]\,. \label{eq:BoltzN}
\end{align}

\subsubsection{Collision Terms for $\phi$}
Here me must track the production of $\phi$ through $L_i\,H\,\rightarrow\, N_{2,3}\,\phi$ and its eventual decay $\phi\, \rightarrow N_i\, N_i$. The freeze-in of $\phi$ takes place at temperatures much higher than the mass of any particle involved, hence we treat all particles as massless, obtaining
\begin{align}
C_{L_i\,H\rightarrow N_j\,\phi}[f_\phi] &= \frac{1}{2E_\phi} \int \diff \Pi_L \diff \Pi_H \diff \Pi_N \M (2\pi)^4 \delta^4\left(\Sigma p\right) f_L(p_L)  f_H(p_H)\nonumber \\
&=\frac{y_{ij}^2}{4\pi^3 M_*^2} ~T^3 \exp(-p_\phi/T)  \,.
\end{align}

The decay process, $\phi \rightarrow N_i \, N_i$ for $i=1,2,3$, gives the corresponding collision term for $f_\phi$:
\begin{align}
C_{\phi \rightarrow N_i \, N_i}[f_\phi] &= \frac{-1}{2E_\phi} \int \diff\Pi_N \diff\Pi_N' \M (2\pi)^4 \delta^4\left(\Sigma p\right) f_\phi(p_\phi) \nonumber \\
&= -\frac{x_{i \text{ eff}}^2\, m_\phi^2}{16\pi E_\phi} f_\phi(p_\phi)\,.
\end{align}

\subsubsection{Collision Terms for $N_1$}

Since dark matter production through $\phi$ decay mainly takes place at temperatures below $m_\phi$, for calculating $f_{N_1}$ we make the approximation that the decaying $\phi$ is at rest:
\be
f_\phi\left(p_\phi, T\right) \simeq 2 \pi^2 n_\phi(T) \frac{\delta(p_\phi)}{p_\phi^2}\,,~~\text{for}~ T\ll m_\phi\,,
\label{eq:phiatrest}
\ee
where $n_\phi(T)$ is determined by solving Eq.\,\ref{eq:BoltzPhi} for $f_\phi(p_\phi,T)$ and integrating over the phase space of $p_\phi$. Inserting this approximation in Eq.\,\ref{eq:phidecaysimplified}, we find
\be
C_{\phi\rightarrow N_1\,N_1}[f_{N_1}] \simeq \frac{\pi x_{1 \text{ eff}}^2}{2 m_\phi} n_\phi(T) \delta\left(p_N - \frac{m_\phi}{2}\right)\,.
\label{eq:s2phidecayatrest}
\ee

\subsection{Scenario III: $\phi$ in equilibrium, supersymmetry}
\label{sec:s3appendix}

Since $\phi$ and $\psi$ are in equilibrium whereas $\tilde{N}_i$ and $N_i$ have negligible abundance at high temperatures, we approximate
\begin{align}
\Omega(\phi\,\phi\rightarrow N_1\,N_1) &\simeq f_\phi f_\phi\,,~~~\Omega(\phi\,\psi\rightarrow N_1\,\tilde{N}_1) \simeq f_\phi f_\psi\,,~~~\Omega(\psi\,\psi\rightarrow \tilde{N}_1\,\tilde{N}_1) \simeq f_\psi f_\psi\,.\nonumber\\
\Omega(\phi\rightarrow N_1\,N_1) &\simeq \Omega(\phi\rightarrow \tilde{N}_1\,\tilde{N}_1) \simeq f_\phi\,,~~~ \Omega(\tilde{N}_1 \rightarrow \psi \, N_1) \simeq f_{\tilde{N}_1}\,.
\end{align}

Note that the phase space densities of $N_1$ and $\tilde{N}_1$ from UV freeze-in should be identical, hence they do not need to be tracked separately, whereas the IR components will differ due to $\phi\rightarrow \tilde{N}_1\,\tilde{N}_1$ proceeding via the soft term.

The Boltzmann equations describing the evolution of $N_1$ and $\Nt$ distributions are
\begin{align}
H r \frac{\partial f_{N_1}}{\partial r}& =  C_{\phi\,\phi\rightarrow N_1\,N_1}[f_{N_1}] + C_{\phi\,\psi\rightarrow N_1\,\tilde{N}_1}[f_{N_1}] + C_{\Nt \rightarrow N_1\,\psi}[f_{N_1}] + C_{\phi\rightarrow N_1\,N_1}[f_{N_1}]\,,\\
H r \frac{\partial f_{\Nt}}{\partial r} &=  C_{\psi\,\psi\rightarrow \Nt\,\Nt}[f_{\Nt}] + C_{\phi\,\psi\rightarrow N_1\,\tilde{N}_1}[f_{\Nt}] + C_{\Nt \rightarrow N_1\,\psi}[f_{\Nt}] + C_{\phi\rightarrow \Nt\,\Nt}[f_{\Nt}]\,.
\end{align}

\subsubsection{UV Freeze-In}

Collision terms describing the UV freeze-in of $N_1$ and $\Nt$ are similar to the $\phi \, \phi \rightarrow N_1 \, N_1$ collision term in Scenario I (Eq.\,\ref{eq:s1annihilation}), similarly resulting in
\begin{align}
C_{\phi\,\phi\rightarrow N_1\,N_1}[f_{N_1}] &= 4\times C_{\phi\,\psi\rightarrow N_1\,\tilde{N}_1}[f_{N_1}] = \frac{4\eta^2}{(2\pi)^3 M_*^2} ~T^3 \exp(-\pn/T) ~\Theta(T-T_d), \\
C_{\psi\,\psi\rightarrow \Nt\,\Nt}[f_{\Nt}] &= 4\times C_{\phi\,\psi\rightarrow N_1\,\tilde{N}_1}[f_{\Nt}] = \frac{4\eta^2}{(2\pi)^3 M_*^2} ~T^3 \exp(-\pNt/T) ~\Theta(T-T_d)\,,
\end{align}
where the factor of 4 accounts for permutations of incoming and outgoing particles.

\subsubsection{IR Freeze-In}

The two collision terms corresponding to the IR freeze-in of $N_1$ are also similar to previously calculated collision terms; we have
\begin{align}
C_{\phi\rightarrow N_1\,N_1}[f_{N_1}] = \frac{ \eeff^2 m_\phi^2}{8 \pi\pn^2} \int_{p_{\phi,\text{min}}}^\infty \diff p_\phi \frac{p_\phi}{E_\phi} f_\phi(p_\phi)\,.
\end{align}
Likewise, since $\Nt$ decays late, we make the approximation that it is at rest at the time of decay
\be
f_{\Nt}\left(p_{\Nt}, T\right) \simeq \pi^2 n_{\Nt}(T) \frac{\delta(p_{\Nt})}{p_{\Nt}^2}\,,~~\text{for}~ T\ll m_{\Nt}\,,
\ee
where the number density $n_{\Nt}$ is found by solving the Boltzmann equations for $\Nt$. The corresponding collision term is
\be
\label{eq:NtdecN}
C_{\Nt\rightarrow N_1\,\psi}[f_{N_1}] \simeq \frac{\pi \eeff^2}{8 m_{\Nt}}\, n_{\Nt}(T)\, \delta\left(p_{N_1} - \frac{m_{\Nt}^2 - m_\psi^2}{2 m_{\Nt}} \right)\,.
\ee

The corresponding collision term for $f_{\Nt}$ can be found in a similar manner to $C_{\phi \rightarrow N_i N_i}[f_\phi]$ in scenario II:
\be
\label{eq:Ntdec}
C_{\Nt\rightarrow N_1\,\psi}[f_{\Nt}] = - \frac{\eeff^2 (m_{\Nt}^2 - m_\psi^2)^2}{16 \pi E_{\Nt} m_{\Nt}^2} f_{\Nt}(p_{\Nt})\,.
\ee

The collision term for $f_{\Nt}$ for $\phi \rightarrow \tilde{N}_i\, \tilde{N}_i$ arising from the soft SUSY breaking term $\eta_i\frac{A_{\eta_i}}{M_*} \phi \phi \tilde{N}_i \tilde{N}_i$ is:
\be
\label{eq:phidecNt}
C_{\phi\rightarrow \Nt\,\Nt}[f_{\Nt}] = \frac{\eeff^2 A_{\eta_i}^2}{4 \pi E_{\Nt} p_{\Nt}} \int_{p_{\phi,\text{min}}}^{p_{\phi,\text{max}}} \diff p_\phi \frac{p_\phi}{E_\phi} f_\phi(p_\phi)\,,
\ee
where
\be
{p_\phi,}_{\renewcommand{\arraystretch}{0.2} \begin{tabular}{l}
  \scriptsize max\\
  \scriptsize min
\end{tabular}} = \frac{\pm\, p_{\Nt}  m_\phi^2 + E_{\Nt} m_\phi \sqrt{m_\phi^2 - 4 m_{\Nt}^2}}{2 m_{\Nt}^2}\,.
\ee

\subsection{Scenario IV: $\phi$ freezes in, supersymmetry}
\label{sec:s4appendix}

The full set of Boltzmann equations describing the evolution of $\phi$, $N_1$, and $\Nt$ are
\begin{align}
H r \frac{\partial f_\phi}{\partial r} &= \sum_{\substack{i=e,\mu,\tau\\ j=2,3}}\Big(C_{L_i\,H\rightarrow N_j\,\phi}[f_\phi] + C_{L_i\,\tilde{H}\rightarrow \tilde{N}_{j}\,\phi}[f_\phi] + C_{\tilde{L}_i\,\tilde{H}\rightarrow N_{j}\,\phi}[f_\phi] \Big)\nonumber \\
& + \sum_{k=1,2,3} \Big( C_{\phi\rightarrow N_k\,N_k}[f_\phi] + C_{\phi\rightarrow \tilde{N}_k\,\tilde{N}_k}[f_\phi] \Big)   \\
H r \frac{\partial f_{N_1}}{\partial r} &= C_{\phi\rightarrow N_1\,N_1}[f_{N_1}] + C_{\Nt \rightarrow N_1\,\psi}[f_{N_1}] \\
H r \frac{\partial f_{\Nt}}{\partial r} &= C_{\Nt \rightarrow N_1\,\psi}[f_{\Nt}] + C_{\phi\rightarrow \Nt\,\Nt}[f_{\Nt}]\,.
\end{align}

\subsubsection{Collision Terms for $\phi$}
Collision terms for $\phi$ are almost identical to those in Scenario II, with additional channels:
\be
\sum_{\substack{i=e,\mu,\tau\\ j=2,3}}\Big(C_{L_i\,H\rightarrow N_j\,\phi}[f_\phi] + C_{L_i\,\tilde{H}\rightarrow \tilde{N}_{j}\,\phi}[f_\phi] +~C_{\tilde{L}_i\,\tilde{H}\rightarrow N_{j}\,\phi}[f_\phi] \Big) = \sum_{\substack{i=e,\mu,\tau\\ j=2,3}} \frac{4 \xi_{ij}^2}{\pi^3 M_*^2} ~T^3 \exp(-p_\phi/T)\,.
\ee
Similarly, for the decay processes,
\begin{align}
C_{\phi \rightarrow N_i \, N_i}[f_\phi] &= -\frac{\eta_{i \text{ eff}}^2\, m_\phi^2}{16\pi E_\phi} f_\phi(p_\phi) \\
C_{\phi \rightarrow \tilde{N}_i \, \tilde{N}_i}[f_\phi] &= -\frac{ \eta_{i \text{ eff}}^2 A_{\eta_i}^2}{8 \pi m_\phi} \sqrt{1-\frac{4 m_{\Nt}^2}{m_\phi^2}}\, \frac{m_\phi}{E_\phi} f_\phi (p_\phi)\,.
\end{align}

\subsubsection{Collision Terms for $\Nt$ and $N_1$}
$\Nt$ production via $\phi \rightarrow \Nt \, \Nt$ occurs when $\phi$ is approximately at rest (see Eq.\,\ref{eq:phiatrest}), giving
\be
C_{\phi\rightarrow \Nt\,\Nt}[f_{\Nt}] \simeq \frac{\pi \eta_{1 \text{ eff}}^2 A_{\eta_1}^2}{4 m_\phi^3 \sqrt{1-4 m_{\Nt}^2/m_\phi^2}} n_\phi(T) \delta\left(p_N - \sqrt{m_\phi^2/4-m_{\Nt}^2}\right) \,.
\ee
Its decay proceeds just as in Scenario III (Eq.\,\ref{eq:Ntdec}).

For decays into $N_1$, as in Scenario II, the decaying $\phi$ and $\Nt$ are taken to be at rest, thus the collision terms for $f_{N_1}$ are identical to Eq.\,\ref{eq:s2phidecayatrest} (for $\phi$ decay) and Eq.\,\ref{eq:NtdecN} (for $\Nt$ decay).

\end{document}